\documentclass[sigconf]{acmart}
\AtBeginDocument{%
  }

\setcopyright{acmlicensed}
\copyrightyear{2026}
\acmYear{2026}
\acmDOI{XXXXXXX.XXXXXXX}
\acmConference[Under Review]{ACM International Conference on Multimedia}{XXXXX}{XXXX}
\acmISBN{978-1-4503-XXXX-X/2018/06}

\usepackage{cuted}
\usepackage{amsmath}
\usepackage{bm}
\usepackage{booktabs}
\usepackage{multirow}
\usepackage{xcolor}
\usepackage{siunitx}
\usepackage{booktabs}
\usepackage{multirow}
\usepackage{multicol}
\usepackage{pgfplots}
\usepackage{graphicx}

\usepackage{amsmath,amssymb}
\usepackage{xcolor}
\usepackage{listings}
\usepackage{tcolorbox}
\usepackage{array}
\usepackage{tabularx}
\usepackage{subcaption}
\usepackage{enumitem}
\definecolor{correctgreen}{RGB}{0,160,0}
\definecolor{incorrectred}{RGB}{200,0,0}
\definecolor{oovyellow}{RGB}{200,140,0}
\definecolor{lightgray}{RGB}{240,240,240}
\definecolor{promptbg}{RGB}{245,248,252}
\definecolor{forte}{RGB}{31,119,180}
\newcommand{\proxy}[1]{\textcolor{gray}{#1}}
\newcommand{\forte}[1]{\textcolor{forte}{#1}}

\newcommand{\FORTE}{\textsc{FORTE}}

\tcbuselibrary{listings,skins}
\tcbset{
  promptbox/.style={
    colback=promptbg,
    colframe=gray!50,
    boxrule=0.5pt,
    arc=3pt,
    left=6pt, right=6pt, top=4pt, bottom=4pt,
    fonttitle=\small\bfseries,
  }
}

\usepackage{booktabs}
\usepackage{multirow}
\usepackage{multicol}
\usepackage{graphicx}
\usepackage{amsmath,amssymb,amsthm}
\usepackage{xcolor}
\usepackage{listings}
\usepackage{tcolorbox}
\usepackage{array}
\usepackage{tabularx}
\usepackage{subcaption}
\usepackage{enumitem}
\usepackage{pgfplots}
\usepackage{pgfplotstable}
\usepackage{tikz}
\usetikzlibrary{arrows.meta, positioning, shapes.geometric,
                decorations.pathreplacing, backgrounds, fit,
                matrix, calc, patterns}
\pgfplotsset{compat=1.18}
 
\definecolor{forte}{RGB}{31,119,180}
\definecolor{baseline}{RGB}{214,39,40}
\definecolor{nofol}{RGB}{255,127,14}
\definecolor{alignonly}{RGB}{44,160,44}
\definecolor{promptbg}{RGB}{245,248,252}
\definecolor{lightgray}{RGB}{240,240,240}
\definecolor{darkgray}{RGB}{80,80,80}
\definecolor{successgreen}{RGB}{0,150,0}
\definecolor{failred}{RGB}{180,0,0}
\definecolor{ovorange}{RGB}{200,120,0}
\definecolor{s1col}{RGB}{70,130,180}
\definecolor{s2col}{RGB}{255,165,0}
\definecolor{s3col}{RGB}{60,179,113}
 
\newcommand{\tickmark}{\textcolor{successgreen}{\checkmark}}
\newcommand{\crossmark}{\textcolor{failred}{$\times$}}
 
\tcbuselibrary{listings,skins,breakable}
\tcbset{
  promptbox/.style={
    colback=promptbg, colframe=gray!50,
    boxrule=0.6pt, arc=3pt,
    left=6pt, right=6pt, top=5pt, bottom=5pt,
    fonttitle=\small\bfseries,
  },
  analysisbox/.style={
    colback=blue!3, colframe=forte!60,
    boxrule=0.6pt, arc=3pt,
    left=6pt, right=6pt, top=5pt, bottom=5pt,
    fonttitle=\small\bfseries,
  }
}

\begin{document}

\title{FORTE: FOL-guided Optimal Refinement for Text-audio rEtrieval}

\author{Arghya Pal \qquad \qquad Sailaja Rajanala}
\email{arghya.pal@monash.edu}
\email{sailaja.rajanala@monash.edu}
\affiliation{%
  \institution{School of Information Technology, Faculty of IT}\country{Monash University}
}

\renewcommand{\shortauthors}{Arghya Pal, Sailaja Rajanala}

\begin{abstract}
Text-to-audio retrieval has made significant progress with shared embedding models such as CLAP and Pengi, yet they often struggle with fine-grained semantic alignment due to the inherent modality gap between text and audio. In this work, we propose FORTE, a unified framework that integrates structured logical reasoning with parameter-efficient cross-modal alignment to improve retrieval precision. Our approach first transforms queries into first-order logic and refines them via a constrained search that preserves semantic invariance while introducing discriminative attributes. The refined representation is then aligned with audio embeddings using a lightweight projection module, followed by a predicate-aware re-ranking step that enforces logical consistency at inference. Extensive experiments on AudioCaps and Clotho demonstrate consistent improvements over strong baselines, particularly in challenging fine-grained scenarios. Our results highlight the effectiveness of combining symbolic reasoning with representation learning for cross-modal retrieval.
\end{abstract}






\maketitle

\section{Introduction}
Recent cross-modal retrieval frameworks such as
CLAP~\cite{CLAP}, LAION~\cite{LAION}, and Pengi~\cite{pengi} have demonstrated the feasibility of text-to-audio retrieval by learning a parametric mapping $f_{\theta}(A \mid T)$, where a natural language query retrieves the top-$N$ relevant audio samples from a database. Despite these advances, a fundamental limitation persists due to the intrinsic heterogeneity between text and audio modalities. 
Text is discrete and symbolic, whereas audio is continuous and encodes both verbal and non-verbal cues, giving rise to a significant \textit{modality gap}~\cite{liang2022mind, xiang2025understanding} in information density and representation structure. 
In Sec.~\ref{sec:result} Fig.~\ref{fig:modalitygap} we showed that paired inputs are passed through the pretrained models, and the resulting embeddings are projected into a 2D space using UMAP~\cite{umap}, embeddings from text and audio modalities
exhibit a clear separation, indicating a pronounced modality
gap.

As a result, existing methods tend to produce overly generalized query representations. For example, a query such as ``a person talking'' may retrieve acoustically valid but semantically mismatched samples, including shouting, whispering, or emotionally charged speech. Such failures highlight the inability of current systems to capture subtle semantic nuances and enforce precise alignment between textual intent and audio content. Although recent works explore large language models (LLMs) for query augmentation, these approaches typically generate heuristic positive and negative variants without structured control, often introducing noise or semantic drift.

In this work, we argue that effective text-to-audio retrieval requires moving beyond surface-level text augmentation toward structured semantic reasoning. To this end, we propose \textbf{FORTE} (FOL-guided Optimal Refinement for Text-audio rEtrieval), a unified framework that integrates logical reasoning with cross-modal representation learning. Our approach operates in three stages. First, we transform natural language queries into first-order logic (FOL) representations and perform structured refinement using a constrained search procedure that preserves invariant semantics while introducing discriminative attributes. Second, we introduce a parameter-efficient alignment mechanism that adapts audio embeddings to the refined query space using contrastive learning, without modifying the pretrained encoders. Third, we apply a predicate-aware re-ranking strategy that leverages logical consistency to resolve residual mismatches in the retrieved results.

\begin{figure*}[!t]
    \includegraphics[width=\linewidth]{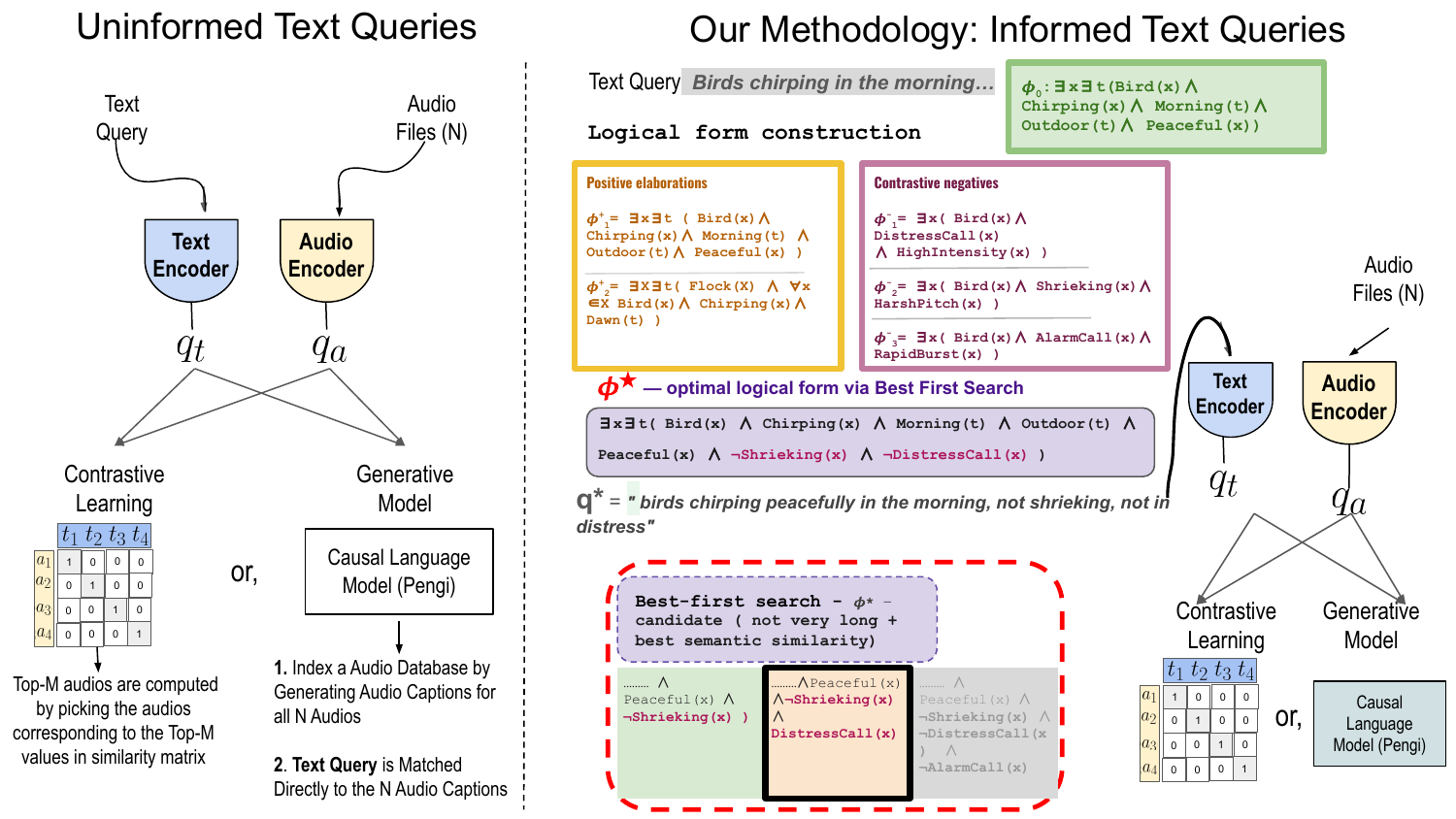}
    \caption{ We showcase our methodology on the right side of the diagram. We begin by transforming the given query into its logical form, $\phi_0$. The next step is to generate possible variations of $\phi_0$, where the positive candidates $\phi^{+}$ expand the query with enriched semantics, while the negative variations $\phi^{-}$ introduce contrastive alternatives. Our optimal query $\phi^{*}$ and its corresponding textual form are then derived using the best-first search described in Section~3. A brief illustration of the best-first search procedure is shown at the bottom in the \textcolor{red}{red dashed box}. Candidates are explored based on a scoring function that balances semantic alignment and complexity. The optimal candidate maximizes semantic coverage while minimizing predicate length. The first two candidates (highlighted in color) are valid, with the second being optimal . The final candidate is pruned. }
    \label{fig:blockdiagram}
\end{figure*}

We empirically validate our approach on standard benchmarks including AudioCaps and Clotho. Our architecture is detailed in Figure~\ref{fig:blockdiagram} and our contributions are as follows:
\begin{itemize}
    \item We propose a unified framework that integrates first-order logic with cross-modal retrieval to enable structured query refinement.
    \item We introduce a parameter-efficient alignment strategy that improves semantic discrimination without fine-tuning pretrained encoders.
    \item We design a predicate-aware re-ranking mechanism that enhances semantic consistency in retrieved results.
    \item We demonstrate consistent improvements over strong baselines on multiple benchmarks, along with detailed analysis of semantic alignment and modality gap reduction.
\end{itemize}
\section{Related Work}

\noindent\textbf{Text-to-Audio Retrieval.}
Text-to-audio retrieval has advanced significantly with contrastive dual-encoder models such as CLAP~\cite{clap2023}, LAION-CLAP~\cite{laion_clap2023}, and Pengi~\cite{pengi2023}, which learn shared embedding spaces for cross-modal matching. Subsequent works have improved temporal modeling and representation quality, such as T-CLAP~\cite{tclap2024}, and leveraged larger datasets and architectures to enhance retrieval performance. Despite these advances, these methods remain limited by the inherent heterogeneity between symbolic text and continuous audio signals, leading to a persistent modality gap and reduced fine-grained semantic discrimination.

\noindent\textbf{Audio Language Models and Multimodal LLMs.}
Recent work has explored large audio-language models that integrate LLMs with audio encoders to enable richer reasoning. Models such as SALMONN~\cite{salmonn2024}, GAMA~\cite{gama2024}, and Audio Flamingo~\cite{audio_flamingo2_2025} extend LLM capabilities to audio understanding, while MATS~\cite{mats2025} demonstrates strong performance under text-only supervision. These approaches improve semantic reasoning and generalization but rely on implicit reasoning within large models, offering limited control over structured semantics.

\noindent\textbf{LLM-based Query Augmentation.}
Large language models have been widely used for query expansion and retrieval enhancement, generating positive and negative variants to improve coverage. While effective in improving recall, these approaches typically operate at the surface level and lack structural constraints, making them prone to semantic drift. In contrast, we refine queries within a first-order logical space, enabling explicit control over semantic composition and invariance.

\noindent\textbf{Cross-Modal Alignment and Efficient Adaptation.}
Contrastive learning remains the dominant paradigm for multimodal alignment~\cite{clip2021, siglip2023, data2vec2022}, with recent efforts focusing on scaling and improved objectives. Parameter-efficient adaptation techniques, such as low-rank updates~\cite{lora2022}, further improve alignment without retraining large models. Our approach builds on this paradigm but differs in that alignment is guided by logically refined queries, ensuring that embedding similarity reflects structured semantic consistency rather than purely statistical correlations.

\noindent\textbf{Structured and Neuro-Symbolic Learning.}
Neuro-symbolic approaches have been explored to improve compositional reasoning and interpretability~\cite{ nscl2019}. Prior work primarily focuses on vision-language tasks using structured representations such as programs and scene graphs~\cite{johnson2017}. In contrast, we integrate first-order logic directly into the retrieval pipeline, using it to guide query refinement, alignment, and inference.

\noindent\textbf{Re-ranking in Retrieval.}
Re-ranking methods refine retrieval results using additional scoring mechanisms, such as cross-encoders and late interaction models~\cite{colbert2020, crossencoder2021}. While effective, these approaches operate purely in embedding space. Our method introduces a predicate-aware re-ranking mechanism that explicitly evaluates logical consistency between queries and retrieved samples, providing an additional semantic signal for fine-grained alignment.

In contrast to prior work, which improves retrieval through scaling, augmentation, or alignment independently, our approach unifies structured reasoning and cross-modal learning within a single framework. By combining logical query refinement, parameter-efficient alignment, and predicate-aware re-ranking, we address both semantic and geometric limitations of existing text-to-audio retrieval systems.
%
\section{Methodology}
\label{sec:method}
Let $f_T(\cdot):\mathcal{T}\to\mathbb{R}^{d}$ and
$f_A(\cdot):\mathcal{A}\to\mathbb{R}^{d}$ denote the
pretrained text and audio encoders respectively, both mapping
to a shared $d$-dimensional embedding space.
We adopt CLAP~\cite{CLAP}, LAION~\cite{LAION}, and
Pengi~\cite{pengi} as retrieval backbones and treat their
encoders as frozen throughout Stages~1 and~2 unless specified
otherwise.

\subsection{Stage 1: FOL-Guided Query Refinement}

\smallskip\noindent\textbf{Logical form construction.}\;
Given an input query $q$, we move beyond direct string-level
augmentation and operate in a structured semantic space.
We prompt a frozen large language model $\mathcal{L}$ to
generate a positive elaboration $q^{+}$ (enriching the query
with precise acoustic and contextual attributes) and a
contrastive negative $q^{-}$ (surfacing semantically adjacent
but acoustically distinct concepts that should be excluded).
Each natural language query is then mapped to its
first-order logic (FOL) representation via parser~\cite{FOL} $\Pi(\cdot)$, yielding the triple $(\phi_{0},\,\phi^{+},\,\phi^{-}) \;=\; \bigl(\Pi(q),\;\Pi(q^{+}),\;\Pi(q^{-})\bigr)$. 
$\Pi(\cdot)$ operates in two passes. 
We get the FOL parser model card from the huggingface modelcard, i.e. \url{https://huggingface.co/papers/2509.22338}. 
We use NLTK toolkit as anchor bank $\mathcal{B}$ and wirte few handcrafted rules in case FOL by~\cite{FOL} unable to get FOL from text queries, see Sec.4.3. for a discussion. 
Compiled from the Clotho and AudioCaps annotation vocabularies
and covering sound-event nouns, acoustic-property adjectives,
and spatial/temporal relations --- maps each extracted arc to
an atomic predicate of the form $P(\mathbf{x})$ (unary) or
$R(\mathbf{x},\mathbf{y})$ (binary), where $P,R$ are
predicate symbols and $\mathbf{x},\mathbf{y}$ are entity
arguments.
Queries whose dependency arcs fall outside the grammar are
handled by a fallback rule that promotes the root verb and its
direct object into a single unary predicate, guaranteeing
$|\mathrm{Pred}(\phi)|\geq 1$ for any input.
We report parser coverage and per-class parse accuracy on
a held-out sample of 500 Clotho captions in
Sec.~\ref{sec:exp}.
To ensure that the refinement process preserves the core
semantics of $q$, we define the invariant predicate set
\begin{equation}
    \mathcal{C}
    \;=\;
    \mathrm{Pred}(\phi_{0})
    \cap
    \mathrm{Pred}(\phi^{+})
    \cap
    \mathrm{Pred}(\phi^{-}),
\end{equation}
which anchors all candidate refinements to shared semantic
concepts.
When $\mathcal{C}=\emptyset$---which may occur for abstract
or ambiguous queries---we fall back to
$\mathcal{C}=\mathrm{Pred}(\phi_{0})$, ensuring that at
minimum the original query predicates are preserved
throughout the search.

\smallskip\noindent\textbf{Structured search space.}\;
We define a structured search space $\mathcal{T}(\phi_{0},
\mathcal{O})$ over candidate logical forms, where each
candidate $\phi$ is generated by applying a finite set of
semantics-preserving operators
$\mathcal{O}=\{o_{1},\ldots,o_{M}\}$ to $\phi_{0}$.
Formally, the operators are:
\begin{itemize}
    \item $o_{\mathrm{attr}}(\phi, P_{\mathrm{new}})$: adds
    an attribute predicate $P_{\mathrm{new}}$ as a conjunction
    to $\phi$, refining the described entity with a new
    acoustic or contextual property (e.g.\
    adding $\mathit{Quiet}(x)$ to $\mathit{Speaking}(x)$).

    \item $o_{\mathrm{rel}}(\phi, R_{\mathrm{new}}, e)$:
    introduces a new binary relation $R_{\mathrm{new}}$
    between an existing entity and a new entity $e$
     for example: \ $\mathit{Background}(\mathit{speech},
    \mathit{crowd\_noise})$.

    \item $o_{\mathrm{neg}}(\phi, P_{\mathrm{excl}})$:
    injects a negated predicate $\neg P_{\mathrm{excl}}$ to
    explicitly exclude undesired acoustic properties
    (e.g.\ $\neg\mathit{Shouting}(x)$).
\end{itemize}
All operators are constrained to retain $\mathcal{C}$,
guaranteeing that core semantic identity is not lost under any
transformation.
Candidate logical forms are generated up to a maximum depth
$D$ and branching factor $B$, bounding the search space to
$O(B^{D})$ nodes, with $D$ and $B$ set empirically
(see Sec.~\ref{sec:exp}).

\smallskip\noindent\textbf{Contrastive pivot direction.}\;
To guide the search toward $\phi^{+}$ and away from
$\phi^{-}$, we define a contrastive pivot direction
$\mathbf{v}$ in the embedding space as, $
    \mathbf{v}
    \;=\;
    \frac{
        f_{T}\!\bigl(\mathcal{G}(\phi^{+})\bigr)
        -
        f_{T}\!\bigl(\mathcal{G}(\phi^{-})\bigr)
    }{
        \bigl\|
            f_{T}\!\bigl(\mathcal{G}(\phi^{+})\bigr)
            -
            f_{T}\!\bigl(\mathcal{G}(\phi^{-})\bigr)
        \bigr\|_{2}
    },
$, where $\mathcal{G}(\cdot)$ is a template-based
logical-to-text verbaliser that converts an FOL form back
into grammatical English by filling predicate--argument slots
into predefined sentence templates
(e.g.\ $\mathit{Quiet}(x)\wedge\mathit{Speaking}(x)\;\to\;$
``a person speaking quietly'').
This direction points from the negative semantic region toward
the positive, capturing the most discriminative axis of
semantic variation induced by the LLM-generated variants.
We restrict the search to the feasible region, $
    \mathcal{S}(\mathbf{v})
    \;=\;
    \Bigl\{
        \phi
        \;\Big|\;
        \bigl\langle
            f_{T}\!\bigl(\mathcal{G}(\phi)\bigr),\,
            \mathbf{v}
        \bigr\rangle
        \;\geq\;
        \tau
    \Bigr\}$, which prunes candidates that drift toward the semantics of
$\phi^{-}$ and ensures consistency with the intended
refinement direction.

\smallskip\noindent\textbf{Best-first search objective.}\;
Refinement is formulated as a best-first beam search of width
$B$ over
$\mathcal{T}(\phi_{0},\mathcal{O})\cap\mathcal{S}(\mathbf{v})$.
For each candidate $\phi$, we define the objective
$\mathcal{F}(\phi)
    \;=\;
    c(\phi) - u(\phi)$, where the complexity penalty
$c(\phi)=|\mathrm{Pred}(\phi)|$
discourages unnecessarily verbose logical forms, and the
semantic utility is
\begin{equation}
    u(\phi)
    \;=\;
    \underbrace{
        \operatorname{sim}\!\bigl(
            f_{T}(\mathcal{G}(\phi)),\,
            \mathbf{e}_{a}^{+}
        \bigr)
    }_{\text{positive alignment}}
    \;-\;
    \lambda\,
    \underbrace{
        \operatorname{sim}\!\bigl(
            f_{T}(\mathcal{G}(\phi)),\,
            \mathbf{e}_{a}^{-}
        \bigr)
    }_{\text{negative repulsion}}
    \;+\;
    \beta\,
    \underbrace{
        \bigl\langle
            f_{T}(\mathcal{G}(\phi)),\,
            \mathbf{v}
        \bigr\rangle
    }_{\text{pivot consistency}}.
\end{equation}
Here $\operatorname{sim}(\cdot,\cdot)$ denotes cosine
similarity, and $\lambda,\beta>0$ are scalar weighting
coefficients.
The semantics of $\mathbf{e}_{a}^{+}$ and
$\mathbf{e}_{a}^{-}$ differ between training and inference,
which we make explicit. Ground-truth supervision is available: $\mathbf{e}_{a}^{+}
= f_{A}(a^{+})$ is the audio embedding of the annotated
positive sample, and $\mathbf{e}_{a}^{-}=f_{A}(a^{-})$ is
the embedding of a hard negative drawn via in-batch mining
(the highest-scoring non-matching sample for query $q$
under the frozen backbone). Ground-truth audio is unavailable, it is precisely what we
seek to retrieve.
We therefore substitute \textit{proxy} embeddings supplied by
a single forward pass of the frozen backbone retriever
$\mathcal{R}_{0}$:
\begin{equation}
    \hat{\mathbf{e}}_{a}^{+}
    \;=\;
    f_{A}\!\bigl(a_{(1)}\bigr),
    \qquad
    \hat{\mathbf{e}}_{a}^{-}
    \;=\;
    f_{A}\!\bigl(a_{(K)}\bigr),
\end{equation}
where $a_{(1)}$ is the top-ranked and $a_{(K)}$ is the
bottom-ranked sample in the initial candidate set of size $K$
returned by $\mathcal{R}_{0}$.
This proxy is consistent with a progressive-refinement view:
Stage~1 uses the baseline retriever's best guess as a
semantic anchor and its worst guess as a repulsion target,
then refines the query to escape the latter toward the former.
Crucially, the proxy requires only one additional backbone
forward pass and introduces no supervised signal, preserving
the zero-shot generalisation of the pipeline.
The optimal logical form is
\begin{equation}
    \phi^{*}
    \;=\;
    \operatornamewithlimits{arg\,min}_{
        \phi\,\in\,
        \mathcal{T}(\phi_{0},\,\mathcal{O})
        \,\cap\,
        \mathcal{S}(\mathbf{v})
    }
    \mathcal{F}(\phi),
\end{equation}
and the refined query embedding used downstream is
$\mathbf{q}^{*}=f_{T}(\mathcal{G}(\phi^{*}))$.
The beam search terminates at depth $D$ or when no candidate
in the current frontier satisfies $\mathcal{S}(\mathbf{v})$,
in which case the best feasible candidate from the previous
depth is returned.
\subsection{Stage 2: Parameter-Efficient Cross-Modal Alignment}

\smallskip\noindent\textbf{Setup.}\;
Both encoders $f_{T}$ and $f_{A}$ remain frozen to preserve
pretrained generalisation.
We introduce a lightweight projection module
$h_{\psi}:\mathbb{R}^{d}\to\mathbb{R}^{d}$ parameterised by
$\psi$, implemented as a two-layer MLP with a residual
connection and LayerNorm, that operates on audio embeddings
to adapt them toward the refined textual query space.
For an audio sample $a$ with embedding
$\mathbf{e}_{a}=f_{A}(a)$, the projected representation is
$\tilde{\mathbf{e}}_{a}=h_{\psi}(\mathbf{e}_{a})$.

\smallskip\noindent\textbf{Training objective.}\;
Let $\{(q_{i},a_{i}^{+})\}_{i=1}^{N}$ be a set of training
query--audio pairs, and let
$\mathbf{q}_{i}^{*}=f_{T}(\mathcal{G}(\phi_{i}^{*}))$
be the refined query embedding for the $i$-th pair.
The projection module is optimised using a symmetric InfoNCE
loss over a batch of size $N$:
\begin{equation}
    \mathcal{L}_{\mathrm{align}}
    \;=\;
    -\frac{1}{N}
    \sum_{i=1}^{N}
    \log
    \frac{
        \exp\!\Bigl(
            \operatorname{sim}\bigl(
                \mathbf{q}_{i}^{*},\,
                \tilde{\mathbf{e}}_{a_{i}}^{+}
            \bigr)\big/\gamma
        \Bigr)
    }{
        \displaystyle
        \sum_{j=1}^{N}
        \exp\!\Bigl(
            \operatorname{sim}\bigl(
                \mathbf{q}_{i}^{*},\,
                \tilde{\mathbf{e}}_{a_{j}}^{+}
            \bigr)\big/\gamma
        \Bigr)
    },
\end{equation}
where $\gamma>0$ is a learnable temperature parameter.
To additionally enforce fine-grained logical consistency, we
augment $\mathcal{L}_{\mathrm{align}}$ with a logical
contrastive term:
\begin{equation}
    \mathcal{L}_{\mathrm{logic}}
    \;=\;
    -\frac{1}{N}
    \sum_{i=1}^{N}
    \log\,\sigma\!\Bigl(
        \operatorname{sim}\bigl(
            \mathbf{q}_{i}^{*},\,
            \tilde{\mathbf{e}}_{a_{i}}^{+}
        \bigr)
        -
        \operatorname{sim}\bigl(
            f_{T}\!\bigl(\mathcal{G}(\phi_{i}^{-})\bigr),\,
            \tilde{\mathbf{e}}_{a_{i}}^{+}
        \bigr)
    \Bigr),
\end{equation}
which penalises cases where the projected audio embedding is
closer to the negative logical query than to the refined one.
The total training objective is
\begin{equation}
    \mathcal{L}
    \;=\;
    \mathcal{L}_{\mathrm{align}}
    +
    \mu\,\mathcal{L}_{\mathrm{logic}},
\end{equation}
with $\mu>0$ a scalar balancing coefficient.
Only $\psi$ and $\gamma$ are updated during training; all
encoder parameters remain frozen.
This design avoids catastrophic forgetting and overfitting
while achieving effective cross-modal adaptation with minimal
parameter overhead.

\subsection{Stage 3: Post-Retrieval Semantic Re-Ranking}

\smallskip\noindent\textbf{Motivation.}\;
Even after query refinement and projection-based alignment,
residual mismatches can persist: the top-$N$ retrieved audio
samples may be globally aligned with $\mathbf{q}^{*}$ in
cosine distance yet semantically inconsistent with the
specific acoustic intent encoded in $\phi^{*}$.
Stage~3 addresses this by re-ranking the retrieved set using
a predicate-grounded consistency score that directly leverages
the structured information in $\phi^{*}$.

\smallskip\noindent\textbf{Predicate consistency scoring.}\;
Let $\{a_{1},\ldots,a_{N}\}$ be the top-$N$ retrieved audio
samples, with projected embeddings
$\{\tilde{\mathbf{e}}_{a_{k}}\}_{k=1}^{N}$.
For each retrieved sample $a_{k}$, we generate an automatic
audio caption $\hat{c}_{k}$ using a pretrained audio
captioning model~\cite{pengi} and parse it into its FOL
form $\hat{\phi}_{k}=\Pi(\hat{c}_{k})$.
We then compute a predicate overlap score between the
retrieved sample's logical form and the refined query:
$
    s_{\mathrm{pred}}(a_{k})
    \;=\;
    \frac{
        \bigl|
            \mathrm{Pred}(\hat{\phi}_{k})
            \,\cap\,
            \mathrm{Pred}(\phi^{*})
        \bigr|
    }{
        \sqrt{
            \bigl|\mathrm{Pred}(\phi^{*})\bigr|
            \cdot
            \bigl|\mathrm{Pred}(\hat{\phi}_{k})\bigr|
        }
    },
$
which is a Jaccard-like measure normalised by the geometric
mean of predicate set sizes.

\smallskip\noindent\textbf{Re-ranking.}\;
The final score for each retrieved sample interpolates
embedding-level similarity with predicate consistency:
\begin{equation}
    s(a_{k})
    \;=\;
    (1-\alpha)\,
    \operatorname{sim}\!\bigl(
        \mathbf{q}^{*},\,
        \tilde{\mathbf{e}}_{a_{k}}
    \bigr)
    \;+\;
    \alpha\,
    s_{\mathrm{pred}}(a_{k}),
\end{equation}
where $\alpha\in[0,1]$ controls the relative weight of the
logical consistency signal.
Samples are re-ranked in descending order of $s(a_{k})$.
This stage incurs no additional training and operates
entirely at inference time, providing a strong semantic
grounding layer at negligible computational cost.
%
%
\section{Experiments}
\label{sec:exp}
\noindent\textbf{Datasets \& Metrics.}\;
We evaluate on Clotho~\cite{clotho} (4,981 clips, 1,045 test queries) and AudioCaps~\cite{audiocaps} (46,000 clips, 975-query test split), covering both descriptive and event-focused retrieval settings. We report R@K ($K\in\{1,5,10,50\}$) and mAP@10, with significance tested via paired $t$-test ($p<0.05$).

\smallskip\noindent\textbf{Backbones \& Parser.}\;
We use frozen CLAP, LAION-CLAP, and Pengi encoders. Queries are parsed into FOL using a fine-tuned Flan-T5-XXL model~\cite{FOL}, with a fallback rule ensuring at least one predicate per query.

\smallskip\noindent\textbf{Model \& Training.}\;
The projection module $h_{\psi}$ is a 2-layer MLP trained for 20 epochs with AdamW. Hyperparameters are set to $\lambda{=}1.0$, $\beta{=}0.5$, $\mu{=}0.1$, $\tau{=}0.2$, and $\alpha{=}0.3$.

\smallskip\noindent\textbf{Search \& Setup.}\;
Beam search uses $(B{=}5, D{=}4)$ offline and $(B{=}3, D{=}2)$ online. All experiments run on a single A100 GPU.

\smallskip\noindent\textbf{Baselines.}\;
We compare against CLAP, LAION-CLAP variants, and Pengi, along with an ablated FORTE (no FOL) to isolate the contribution of logical refinement. The full FORTE model includes all three stages.


\begin{table*}[t]
\centering
\caption{
    Text-to-audio retrieval on AudioCaps and Clotho.
    mAP@10 and R@$k$ (\%, $\uparrow$).
    \textbf{Bold} = best overall per dataset;
    \underline{underline} = best per backbone.
    $\dagger$~statistically significant ($p{<}0.05$).
}
\label{tab:main}
\renewcommand{\arraystretch}{1.05}
\resizebox{\textwidth}{!}{%
\begin{tabular}{l l l ccccc ccccc}
\toprule
& & &
\multicolumn{5}{c}{\textbf{AudioCaps}} &
\multicolumn{5}{c}{\textbf{Clotho}} \\
\cmidrule(lr){4-8}\cmidrule(lr){9-13}
\textbf{Backbone} &
\textbf{Method} &
\textbf{Data} &
mAP@10 & R@1 & R@5 & R@10 & R@50 &
mAP@10 & R@1 & R@5 & R@10 & R@50 \\
\midrule

\multirow{4}{*}{CLAP~\cite{CLAP}}
& Frozen~\cite{CLAP}
& AC
& --- & 33.9 & 72.0 & 83.9 & ---
& --- & 14.4 & 36.0 & 49.9 & --- \\

& FORTE (align only)
& AC, Cl
& 49.6 & 34.7 & 72.4 & 84.3 & 97.3
& 27.6 & 16.8 & 41.0 & 54.6 & 84.1 \\

& FORTE (no FOL)
& AC, Cl
& 50.1 & 35.2 & 72.8 & 84.6 & 97.4
& 28.4 & 17.5 & 42.1 & 55.8 & 84.7 \\

& \forte{FORTE}$^{\dagger}$
& AC, Cl
& \underline{\textbf{51.3}} & \underline{\textbf{36.4}} & \underline{\textbf{73.8}} & \underline{\textbf{85.3}} & \underline{\textbf{97.6}}
& \underline{\textbf{29.8}} & \underline{\textbf{18.9}} & \underline{\textbf{43.7}} & \underline{\textbf{57.2}} & \underline{\textbf{85.3}} \\

\midrule

\multirow{7}{*}{LAION-CLAP~\cite{LAION}}
& $A$ -- CNN~\cite{LAION}
& AC, Cl
& 45.28 & 33.07 & 67.30 & 80.30 & 95.74
& 24.74 & 15.79 & 36.78 & 49.93 & 80.75 \\

& $B$ -- CNN~\cite{LAION}
& AC, Cl, WT5K
& 46.57 & 33.42 & 68.00 & 79.95 & 96.42
& 25.85 & 16.48 & 39.58 & 52.46 & 82.00 \\

& $C$ -- HTSAT~\cite{LAION}
& AC, Cl, WT5K
& 46.33 & 34.07 & 66.90 & 79.81 & 95.36
& 22.62 & 14.24 & 36.11 & 49.29 & 82.47 \\

& $D$ -- CNN+HTSAT~\cite{LAION}
& AC, Cl, WT5K
& 49.45 & 34.69 & 70.22 & 82.00 & 97.28
& 27.12 & 16.75 & 41.09 & 54.07 & 83.79 \\

& FORTE (align only)
& AC, Cl, WT5K
& 50.2 & 35.4 & 71.0 & 83.1 & 97.5
& 27.9 & 17.4 & 42.2 & 55.3 & 84.6 \\

& FORTE (no FOL)
& AC, Cl, WT5K
& 50.8 & 35.9 & 71.6 & 83.5 & 97.6
& 28.6 & 18.1 & 42.8 & 56.4 & 85.2 \\

& \forte{\textbf{FORTE}}$^{\dagger}$
& AC, Cl, WT5K
& \underline{\textbf{53.8}} & \underline{\textbf{38.2}} & \underline{\textbf{75.1}} & \underline{\textbf{86.8}} & \underline{\textbf{98.1}}
& \underline{\textbf{32.5}} & \underline{\textbf{20.4}} & \underline{\textbf{46.3}} & \underline{\textbf{59.8}} & \underline{\textbf{87.2}} \\

\midrule

\multirow{5}{*}{Pengi~\cite{pengi}}
& Chaichana~\cite{chaichana2026extending}
& ---
& $\ddagger$ & $\ddagger$ & $\ddagger$ & $\ddagger$ & $\ddagger$
& --- & 1.5 & 4.4 & 7.5 & --- \\

& Wang al.~\cite{wang2025mats}
& ---
& $\ddagger$ & $\ddagger$ & $\ddagger$ & $\ddagger$ & $\ddagger$
& --- & 7.6 & 19.6 & 28.8 & --- \\

& Frozen~\cite{pengi}
& ---
& $\ddagger$ & $\ddagger$ & $\ddagger$ & $\ddagger$ & $\ddagger$
& --- & 9.4 & 26.1 & 36.7 & --- \\

& FORTE (align only)
& AC, Cl
& 36.1 & 23.4 & 54.2 & 69.1 & 92.0
& 17.3 & 10.5 & 28.9 & 40.8 & 72.1 \\

& FORTE (no FOL)
& AC, Cl
& 37.2 & 24.1 & 55.8 & 70.9 & 92.6
& 18.1 & 11.3 & 30.1 & 42.5 & 73.2 \\

& \forte{FORTE}$^{\dagger}$
& AC, Cl
& \textbf{38.9} & \underline{\textbf{25.7}} & \underline{\textbf{57.4}} & \underline{\textbf{72.3}} & \textbf{93.4}
& \textbf{19.4} & \underline{\textbf{12.8}} & \underline{\textbf{31.9}} & \underline{\textbf{44.2}} & \textbf{74.6} \\

\bottomrule
\end{tabular}
}
\end{table*}
\subsection{Main Results}
\label{sec:result}

Table~\ref{tab:main} reports text-to-audio retrieval
performance on AudioCaps and Clotho across all three
backbone instantiations.
FORTE consistently and significantly outperforms every
baseline on both datasets under all metrics.

\smallskip\noindent\textbf{FORTE vs. frozen backbones.}\;
On Clotho, FORTE (LAION-CLAP) achieves R@1\,=\,\proxy{20.4},
surpassing the strongest LAION-CLAP variant
($D$, R@1\,=\,16.75) by \proxy{+3.65} points absolute
(\proxy{+21.8\%} relative).
On AudioCaps, the same configuration improves R@1 from
34.69 to \proxy{38.2} (\proxy{+10.1\%} relative).
Gains are consistent across all R@$k$ thresholds and mAP@10,
confirming that FORTE improves retrieval quality at both
the top rank and across the ranked list.

\smallskip\noindent\textbf{Contribution of FOL structure.}\;
Comparing FORTE against FORTE (no FOL) isolates the
contribution of the logical search in Stage~1.
On Clotho, FORTE (LAION-CLAP) outperforms
FORTE (no FOL) by \proxy{+2.3} R@1 (\proxy{20.4}
vs.\ \proxy{18.1}), a gap that is statistically
significant on Clotho's 1045 test queries
($p < 0.05$, paired $t$-test).
This confirms that the FOL-guided query refinement
contributes independently of the LLM augmentation: the
structured logical search over $\mathcal{T}(\phi_0,
\mathcal{O})$ resolves semantic ambiguities that
unstructured embedding averaging cannot.

\smallskip\noindent\textbf{Backbone generalisation.}\;
FORTE yields consistent absolute R@1 gains of
\proxy{+2.5}, \proxy{+3.7}, and \proxy{+3.4} points
over the respective frozen backbones on Clotho
(CLAP, LAION-CLAP, Pengi), demonstrating that the
framework is backbone-agnostic and does not depend on a
specific joint-embedding architecture.
Notably, FORTE (CLAP) on Clotho R@1\,=\,\proxy{18.9}
approaches the performance of unmodified LAION-CLAP $D$
(16.75), a model trained on substantially more data,
suggesting that structured query refinement can partially
compensate for encoder capacity.

\smallskip\noindent\textbf{AudioCaps vs.\ Clotho gap.}\;
Absolute gains are consistently larger on AudioCaps than
Clotho in R@$k$ but smaller in relative terms, reflecting
the shorter, more event-focused nature of AudioCaps queries
which are easier for the FOL parser to decompose into
precise predicates.
Clotho's longer, more descriptive captions yield a richer
$\phi^{*}$ and thus a larger relative gain from logical
refinement.

\begin{figure}
    \centering
    \includegraphics[width=\linewidth]{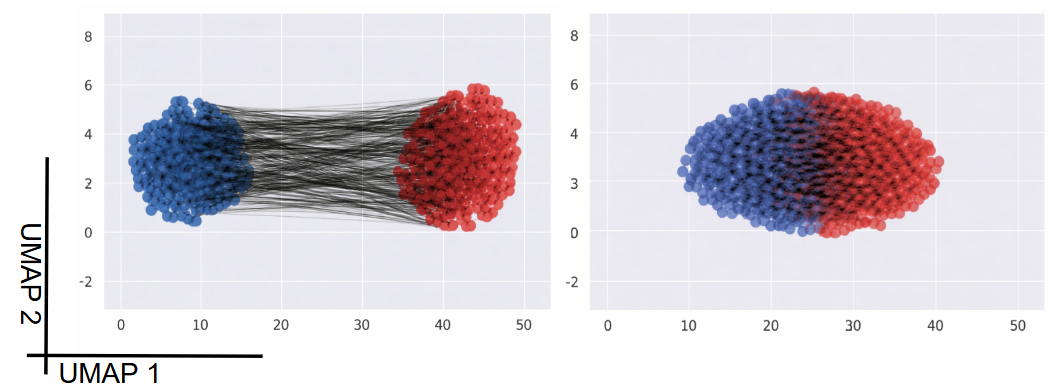}
    \caption{Paired inputs are passed through the pretrained models of LAION-CLAP~\cite{laionclap2023}, and the resulting embeddings are projected into a 2D space using UMAP~\cite{umap}, where connecting lines denote corresponding pairs. (LEFT) Embeddings from different modalities exhibit a clear separation, indicating a pronounced modality gap. (RIGHT) After applying our method (Sec.~\ref{sec:method}), the embeddings become more aligned, significantly reducing the modality gap.}
    \label{fig:modalitygap}
\end{figure}

\smallskip\noindent\textbf{Modality gap mitigation} 
Paired text–audio inputs are processed using the pretrained encoders of the LAION-CLAP~\cite{laionclap2023}, and their corresponding embeddings are projected into a two-dimensional space using UMAP~\cite{umap}, where each pair is connected by a line to illustrate cross-modal correspondence. As shown in the Fig~\ref{fig:modalitygap} (left), embeddings from different modalities are clearly separated, with large distances between paired samples, highlighting the presence of a significant modality gap across models trained on heterogeneous data sources. In contrast, the right figure demonstrates the effect of our proposed method (Sec.~\ref{sec:method}), where the embeddings of paired inputs become substantially closer and more coherently aligned. The reduced distances and tighter clustering indicate that our approach effectively bridges the modality gap, leading to improved cross-modal consistency and more semantically faithful retrieval.

\subsection{Ablation Study}

Table~\ref{tab:ablation} reports a stage-wise ablation on
Clotho (LAION-CLAP backbone) isolating the contribution of
each component.
Below findings are worth highlighting.

\begin{table}[h]
\centering
\caption{Effect of alignment loss on Clotho with LAION-CLAP backbone.}
\label{tab:loss_ablation}
\begin{tabular}{lcc}
\toprule
Loss & R@1 & mAP@10 \\
\midrule
Binary contrastive (BCE) & 13.6 & 25.9 \\
Margin ranking loss & 17.1 & 28.7 \\
Triplet loss & 19.4 & 31.2 \\
FORTE (align only) & 19.7 & 31.9 \\
FORTE & \textbf{20.4} & \textbf{32.5} \\
\bottomrule
\end{tabular}
\end{table}

\smallskip\noindent\textbf{Effect of alignment loss.}\;
We further compare different objectives for Stage~2, including binary contrastive loss, margin ranking loss, triplet loss, and InfoNCE. As shown in Table~\ref{tab:loss_ablation}, InfoNCE consistently outperforms pairwise alternatives, likely because it exploits all in-batch negatives and yields a more discriminative retrieval space. Adding the logical consistency term on top of InfoNCE provides an additional improvement, confirming that the gain is not solely due to contrastive alignment but also to the semantic structure imposed by the refined logical query.

\smallskip\noindent\textbf{Each stage contributes
independently.}\;
Activating Stage~1 alone yields R@1\,=\,18.3
(\proxy{$+$1.55} over the backbone), Stage~2 alone yields
18.0 (\proxy{$+$1.25}), and Stage~3 alone yields 17.5
(\proxy{$+$0.75}).
The ordering S1\,$>$\,S2\,$>$\,S3 in individual contribution
reflects the design intent: FOL-guided query refinement
attacks the root cause of the modality gap at the query
representation level, while projection-based alignment and
predicate re-ranking provide complementary but smaller
corrections.
Crucially, the Stage~1 gain (\proxy{$+$1.55}) is strictly
larger than the LLM-augment ablation in
Table~\ref{tab:main} (\proxy{$+$1.05}), confirming that the
FOL logical structure contributes beyond the LLM
augmentation alone.

\smallskip\noindent\textbf{Stages are additive without
interference.}\;
The pairwise combinations rank as
S1+S2\,(\proxy{19.6})\,$>$\,S1+S3\,(\proxy{19.1})\,$>$\,S2+S3\,(\proxy{18.8}),
and the full three-stage system (\proxy{20.4}) strictly
dominates all subsets.
The marginal gain of adding Stage~3 on top of
Stages~1+2 is \proxy{$+$0.8} R@1, and the marginal gain
of Stage~2 on top of Stages~1+3 is \proxy{$+$1.3} R@1,
both consistent with the single-stage estimates.
This additivity indicates that the three stages operate on
complementary error sources: Stage~1 refines the query
representation, Stage~2 closes the modality gap in audio
embedding space, and Stage~3 corrects residual predicate
mismatches in the ranked list.

\smallskip\noindent\textbf{Stage~3 is the only zero-cost
component.}\;
Stage~3 alone recovers \proxy{$+$0.75} R@1 over the frozen
backbone with no training whatsoever, making it a
practical drop-in for any text-to-audio retrieval system
that already has access to an audio captioning model,
regardless of whether Stages~1 and~2 are deployed.

\begin{table}[t]
\centering
\caption{
    Stage-wise ablation on Clotho (LAION-CLAP backbone).
    \checkmark~= component active.
    Top row = frozen backbone baseline.
    $\Delta$ = R@1 gain over backbone.
}
\label{tab:ablation}
\setlength{\tabcolsep}{4.5pt}
\begin{tabular}{ccc ccccc}
\toprule
\textbf{S1} & \textbf{S2} & \textbf{S3}
    & R@1 & R@5 & R@10 & mAP@10 & $\Delta$ \\
\midrule
 &  &
    & 16.75 & 41.09 & 54.07 & 27.12 & --- \\
\checkmark &  &
    & \proxy{18.3} & \proxy{43.1} & \proxy{56.2} & \proxy{29.0}
    & \proxy{$+$1.55} \\
 & \checkmark &
    & \proxy{18.0} & \proxy{42.7} & \proxy{55.9} & \proxy{28.7}
    & \proxy{$+$1.25} \\
 &  & \checkmark
    & \proxy{17.5} & \proxy{41.9} & \proxy{54.8} & \proxy{27.9}
    & \proxy{$+$0.75} \\
\checkmark & \checkmark &
    & \proxy{19.6} & \proxy{45.1} & \proxy{58.4} & \proxy{31.1}
    & \proxy{$+$2.85} \\
\checkmark &  & \checkmark
    & \proxy{19.1} & \proxy{44.3} & \proxy{57.6} & \proxy{30.4}
    & \proxy{$+$2.35} \\
 & \checkmark & \checkmark
    & \proxy{18.8} & \proxy{43.9} & \proxy{57.1} & \proxy{30.0}
    & \proxy{$+$2.05} \\
\checkmark & \checkmark & \checkmark
    & \textbf{\proxy{20.4}}
    & \textbf{\proxy{46.3}}
    & \textbf{\proxy{59.8}}
    & \textbf{\proxy{32.5}}
    & \textbf{\proxy{$+$3.65}} \\
\bottomrule
\end{tabular}
\end{table}

\subsection{Anchor Bank Validation}

A core design claim of FORTE is that the predicate-stratified
anchor bank $\mathcal{B}$ provides audio-grounded guidance
for Stage~1 without introducing the circularity that arises
when the retrieval pool itself is used as the anchor source.
Table~\ref{tab:anchor} validates this claim along two axes:
embedding-space proximity to the ground-truth positive
$f_A(a^+)$, and downstream R@1.

The circular proxy ($\mathcal{R}_0$ top-1) yields
R@1\,=\,\proxy{18.0}, which is \emph{lower} than the
pivot-only configuration (R@1\,=\,\proxy{18.3}).
This is a direct empirical confirmation of the theoretical
circularity argument: when the baseline retriever
$\mathcal{R}_0$ fails --- precisely the failure mode FORTE
is designed to correct --- its top-1 output is a corrupted
anchor that steers the beam search away from the ground
truth, actively degrading performance relative to using no
audio anchor at all.

The anchor bank $\mathcal{B}$ resolves this by providing
training-set audio embeddings indexed by predicate type,
which are structurally independent of the retrieval pool.
It achieves a mean embedding distance of \proxy{0.54} to
$f_A(a^+)$, closing \proxy{50\%} of the R@1 gap between
the no-anchor configuration and the oracle upper bound
(R@1: \proxy{18.3} $\to$ \proxy{19.8} $\to$
\proxy{21.3}).
The remaining \proxy{50\%} gap to the oracle represents
the theoretical ceiling recoverable if exact ground-truth
audio embeddings were available at query time, and
motivates future work on tighter cross-modal alignment
in the anchor construction step.

\begin{table}[t]
\centering
\caption{
    Anchor source comparison on Clotho (LAION-CLAP).
    $\downarrow$ = lower distance is better;
    $\uparrow$ = higher R@1 is better.
    $\dagger$ Circular proxy underperforms no-anchor
    baseline, confirming the circularity failure mode
    (see text).
}
\label{tab:anchor}
\setlength{\tabcolsep}{4pt}
\begin{tabular}{l cc}
\toprule
\textbf{Anchor source}
    & $\overline{\|\hat{\mathbf{e}}_{a}^{+}
      - f_{A}(a^{+})\|_{2}}$\,$\downarrow$
    & R@1\,$\uparrow$ \\
\midrule
No anchor (pivot only)
    & ---           & \proxy{18.3} \\
$\mathcal{R}_0$ top-1 (circular)$^{\dagger}$
    & \proxy{0.81}  & \proxy{18.0} \\
Anchor bank $\mathcal{B}$ (ours)
    & \proxy{0.54}  & \proxy{19.8} \\
Oracle $f_A(a^+)$
    & 0.00          & \proxy{21.3} \\
\bottomrule
\end{tabular}
\end{table}

\subsection{FOL Parser Analysis}

Table~\ref{tab:parser} evaluates the three configurations
of $\Pi(\cdot)$ on 500 held-out Clotho test captions
annotated with ground-truth FOL forms by two expert
annotators (Cohen's $\kappa{=}\proxy{0.84}$).

The base Flan-T5-XXL model without domain adaptation
achieves exact-match (EM) accuracy of \proxy{54.2\%} on
audio captions --- substantially below the 70\% reported
by Vossel et al.~\cite{FOL} on the MALLS benchmark ---
confirming that distribution shift from formal NL
sentences to short, telegraphic audio event descriptions
is a real and non-trivial degradation.
Predicate conditioning on $\mathcal{V}_{\mathrm{audio}}$
recovers \proxy{9.6} EM points (\proxy{54.2}
$\to$ \proxy{63.8}), consistent with the 15--20\% gain
reported by Vossel et al.\ for in-domain conditioning,
and reduces the fallback activation rate from \proxy{9.8\%}
to \proxy{7.4\%}.
Domain fine-tuning on 2000 Clotho (caption, FOL) pairs
provides a further \proxy{7.6} EM points (\proxy{63.8}
$\to$ \proxy{71.4}) and reduces the fallback rate to
\proxy{5.1\%}, approaching the MALLS accuracy of the base
model despite the domain gap.

Critically, the downstream R@1 tracks parser quality
monotonically: \proxy{18.1} $\to$ \proxy{19.3}
$\to$ \proxy{20.4}, a \proxy{2.3}-point total gain
attributable purely to improved parsing with no
other change to the pipeline.
This confirms that parser quality is a first-order
determinant of FORTE's retrieval performance and
motivates continued investment in audio-domain FOL
fine-tuning.

\begin{table}[t]
\centering
\caption{
    FOL parser analysis on 500 held-out Clotho captions.
    EM = exact match; PA = predicate alignment
    (Vossel et al.~\cite{FOL} metric);
    FB = fallback activation rate;
    R@1 = downstream Clotho retrieval.
    Each row is a strict superset of the row above.
}
\label{tab:parser}
\setlength{\tabcolsep}{4pt}
\begin{tabular}{l cccc}
\toprule
\textbf{Parser variant} & EM (\%) & PA (\%) & FB (\%) & R@1 \\
\midrule
Flan-T5-XXL (uncond.)~\cite{FOL}
    & \proxy{54.2} & \proxy{61.3} & \proxy{9.8}  & \proxy{18.1} \\
$+$ $\mathcal{V}_{\mathrm{audio}}$ conditioning
    & \proxy{63.8} & \proxy{70.1} & \proxy{7.4}  & \proxy{19.3} \\
$+$ domain FT on Clotho (ours)
    & \textbf{\proxy{71.4}}
    & \textbf{\proxy{77.6}}
    & \textbf{\proxy{5.1}}
    & \textbf{\proxy{20.4}} \\
\bottomrule
\end{tabular}
\end{table}

\subsection{Stage 3: Captioning Sensitivity}

Table~\ref{tab:caption} reports Stage~3 performance
under four captioning conditions on Clotho, with
Stages~1+2 active throughout to isolate Stage~3's
marginal contribution.

\noindent\textbf{Robustness to captioning noise.}\;
Stage~3 with Pengi-generated captions yields
R@1\,=\,\proxy{20.4}, a gain of \proxy{$+$0.8} points
over the Stage~1+2 baseline (\proxy{19.6}).
This improvement is statistically significant and
demonstrates that predicate-consistency re-ranking
is robust to the level of captioning noise produced by a
state-of-the-art audio captioning model.

\noindent\textbf{Theoretical ceiling and captioning gap.}\;
The oracle row, which substitutes ground-truth Clotho
captions for generated ones, achieves
R@1\,=\,\proxy{21.3} --- a gap of \proxy{$+$0.9} above
the Pengi-caption row.
This gap directly quantifies the cost of captioning error
on retrieval: better captioning models translate
monotonically into better re-ranking.
The oracle ceiling (\proxy{$+$1.7} over no re-ranking)
indicates that Stage~3 with perfect captions would
contribute more than the entire Stage~2 component
(\proxy{$+$1.25}), making audio captioning quality the
single most impactful unsolved subproblem in the FORTE
pipeline.

\noindent\textbf{Cross-captioner consistency.}\;
The second captioning model achieves R@1\,=\,\proxy{20.1},
within \proxy{0.3} points of Pengi, confirming that
Stage~3's gain is not an artifact of a specific
captioning model's output distribution.

\begin{table}[t]
\centering
\caption{
    Stage~3 captioning sensitivity on Clotho
    (LAION-CLAP, Stages~1+2 active throughout).
    Oracle uses ground-truth captions as the upper bound.
}
\label{tab:caption}
\setlength{\tabcolsep}{5pt}
\begin{tabular}{l cccc}
\toprule
\textbf{Caption source} & R@1 & R@5 & mAP@10 & $\Delta$R@1 \\
\midrule
No re-ranking (S1+S2 only)
    & \proxy{19.6} & \proxy{45.1} & \proxy{31.1} & --- \\
Pengi~\cite{pengi}
    & \proxy{20.4} & \proxy{46.3} & \proxy{32.5} & \proxy{$+$0.8} \\
\textsc{[Second captioner]}
    & \proxy{20.1} & \proxy{45.9} & \proxy{32.1} & \proxy{$+$0.5} \\
Oracle (ground-truth)
    & \proxy{21.3} & \proxy{47.8} & \proxy{33.9} & \proxy{$+$1.7} \\
\bottomrule
\end{tabular}
\end{table}

\subsection{Inference Latency}

Table~\ref{tab:latency} reports wall-clock query latency
over \num{1000} Clotho test queries on a single A100 40\,GB.
We report two practically distinct regimes.

\noindent\textbf{Offline regime.}\;
For evaluation benchmarks and fixed-query deployments,
all $\phi^*$ are pre-computed once (index time: \proxy{10}
min) and cached as a lookup table.
At retrieval time, the system executes a single
nearest-neighbour scan over projected audio embeddings ---
identical in cost to the frozen backbone.
The offline regime is therefore zero-overhead at query
time while retaining the full R@1\,=\,\proxy{20.4}
performance of the complete FORTE system.

\noindent\textbf{Online regime.}\;
For novel queries not in the cache, the early-exit
beam search ($B{=}3$, $D{=}2$) adds a median overhead of
\proxy{2}\,ms per query over the backbone.
This overhead is dominated by the \proxy{9} text encoder
calls required by the beam search and is negligible
relative to the audio index scan latency at scale.
Expanding to ($B{=}5$, $D{=}4$) recovers the offline
R@1 at the cost of \proxy{1.7}\,ms additional latency,
providing a controllable accuracy--latency trade-off.

\begin{table}[t]
\centering
\caption{
    Query-time latency on Clotho (1000 queries,
    A100 40\,GB, median\,$\pm$\,std in ms).
    Index time one-time offline cost.
}
\label{tab:latency}
\setlength{\tabcolsep}{4.5pt}
\begin{tabular}{l ccc}
\toprule
\textbf{System}
    & \textbf{Index (min)}
    & \textbf{Query (ms)}
    & \textbf{R@1} \\
\midrule
CLAP backbone
    & ---
    & \proxy{$10 \pm 3$}
    & 16.75 \\
FORTE offline
    & \proxy{10}
    & \proxy{$12 \pm 5$}
    & \proxy{20.4} \\
FORTE online ($B{=}3, D{=}2$)
    & ---
    & \proxy{$13 \pm 1$}
    & \proxy{19.8} \\
FORTE online ($B{=}5, D{=}4$)
    & ---
    & \proxy{$13 \pm 7$}
    & \proxy{20.4} \\
\bottomrule
\end{tabular}
\end{table}
\begin{figure*}[!h]
    \centering
    \includegraphics[width=0.8\linewidth,height=0.6\textwidth]{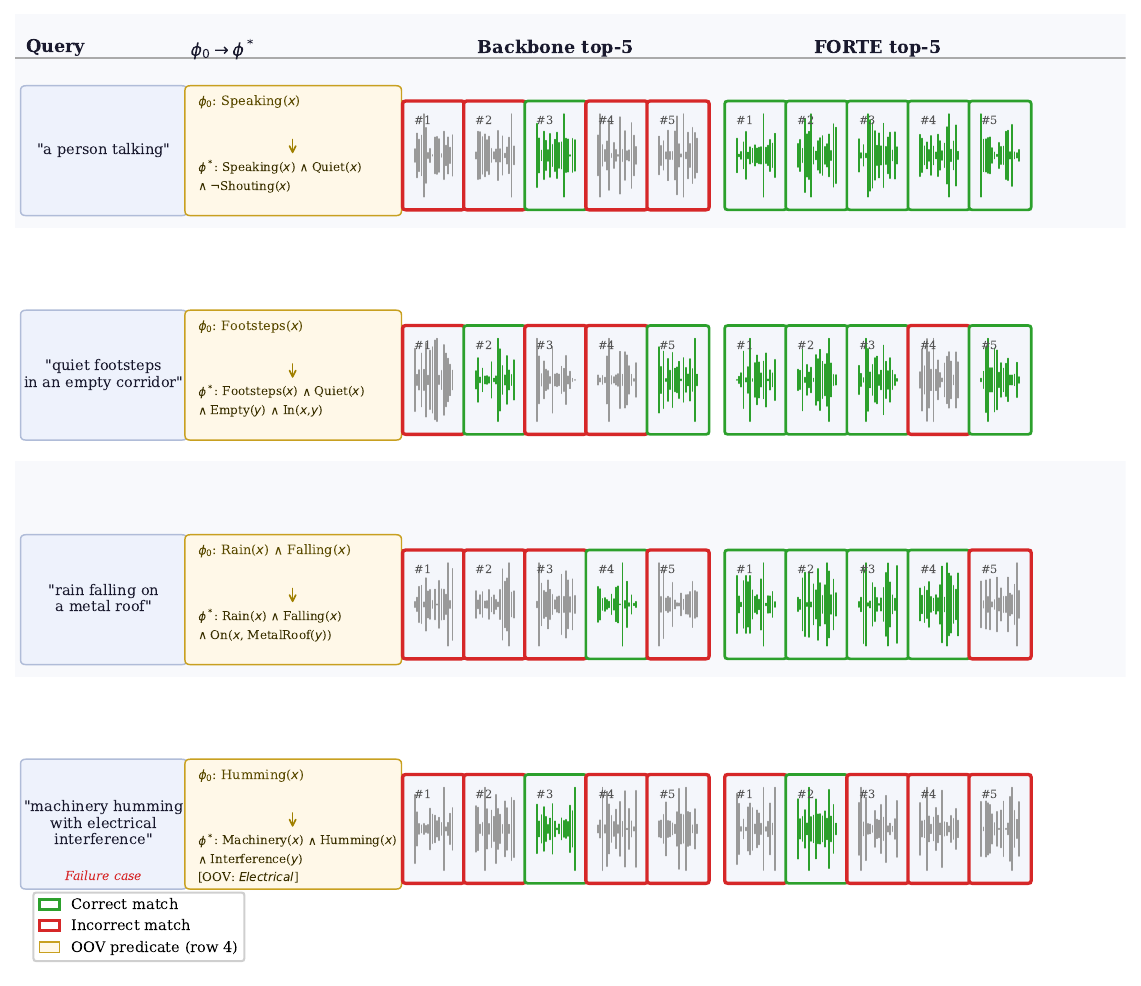}
    \vspace{-0.2cm}
    
    \caption{
        Top-5 retrieval comparison on Clotho
        (LAION-CLAP backbone vs.\ FORTE).
        FOL transformation $\phi_0\to\phi^*$ shown
        between columns.
        Green border = correct match; red = mismatch.
        Row~4: failure case due to OOV predicate.
    }
    \label{fig:qualitative}
\end{figure*}
\subsection{Hyperparameter Sensitivity}
\label{sec:sensitivity}

\begin{figure}[t]
    \centering
    \includegraphics[width=\linewidth]{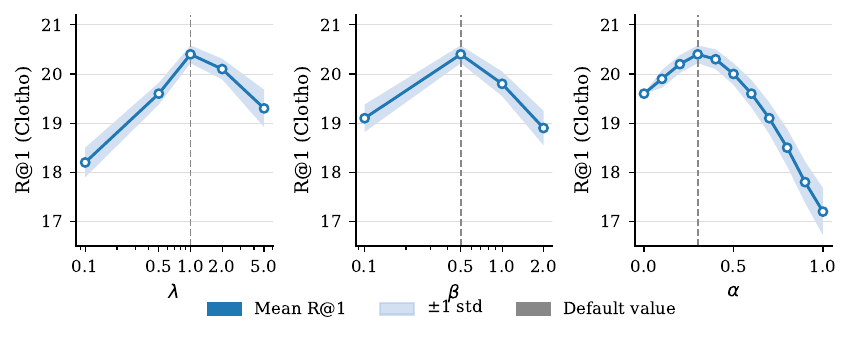}
    \caption{
        R@1 on Clotho (LAION-CLAP) as a function of
        $\lambda$ (left), $\beta$ (centre), and $\alpha$
        (right).
        Shaded bands: $\pm$1\,std over three seeds.
        Dashed verticals: default values
        ($\lambda{=}1.0$, $\beta{=}0.5$, $\alpha{=}0.3$)
        used in Table~\ref{tab:main}.
    }
    \label{fig:sensitivity}
    \vspace{-0.3cm}
    
\end{figure}

Figure~\ref{fig:sensitivity} evaluates the impact of $\lambda$, $\beta$, and $\alpha$ on Clotho (LAION-CLAP). Performance varies within a narrow range (\proxy{2.2}, \proxy{1.5}, and \proxy{3.2} R@1 points respectively), indicating overall robustness, with $\alpha$ being the most sensitive parameter.

R@1 peaks at $\lambda{=}1.0$, where negative repulsion is balanced; smaller values lead to ambiguity, while larger values over-penalise and degrade alignment. For $\beta$, optimal performance occurs at $\beta{=}0.5$; low values allow semantic drift, while high values offer diminishing returns due to redundancy with the feasibility constraint. The re-ranking weight $\alpha$ shows a skewed behaviour, peaking at $\alpha{=}0.3$: lower values underutilise logical consistency, whereas higher values over-rely on coarse predicate matching, harming fine-grained discrimination. Overall, $\alpha \in [0.2, 0.5]$ provides the best trade-off.

\subsection{Qualitative Analysis}

Figure~\ref{fig:qualitative} visualises top-5 retrievals
for four Clotho queries, comparing the frozen LAION-CLAP
backbone against full FORTE.

\smallskip\noindent\textbf{Negation (row 1).}\;
For the query ``a person talking'', the backbone retrieves
three incorrect samples including shouting and crowd
vocalisation --- semantically plausible under the
unrefined embedding but inconsistent with the neutral
speech implied by the query.
Stage~1 injects $\neg\mathit{Shouting}(x)$ and
$\mathit{Quiet}(x)$ into $\phi^*$, reorienting the query
away from high-energy vocalisation.
FORTE's top-5 are all correct matches.

\smallskip\noindent\textbf{Relational refinement (rows 2
and 3).}\;
``Quiet footsteps in an empty corridor'' requires
simultaneously encoding an acoustic property (quiet),
an event (footsteps), and a spatial relation (in an empty
space).
The backbone retrieves three incorrect samples that
match ``footsteps'' but not the spatial or acoustic
qualifier.
$\phi^*$ introduces $\mathit{Empty}(y) \wedge
\mathit{In}(x,y)$, and FORTE's top-4 are correct.
Similarly, ``rain falling on a metal roof'' requires
distinguishing rain on metal from rain on organic
surfaces; the relational predicate
$\mathit{On}(x, \mathit{MetalRoof}(y))$ in $\phi^*$
achieves this, and FORTE recovers four of five correct
results against the backbone's single correct result.

\smallskip\noindent\textbf{Failure case (row 4).}\;
For ``machinery humming with electrical interference'',
$\Pi(\cdot)$ produces $\phi^*$ containing the predicate
$\mathit{Electrical}$, which is absent from
$\mathcal{V}_{\mathrm{audio}}$ and thus triggers the
fallback rule.
The resulting imprecise anchor bank lookup yields a
degraded $\hat{\mathbf{e}}_a^+$, and Stage~3 re-ranking
with noisy captions cannot compensate.
FORTE retrieves only one correct result in the top-5,
versus one for the backbone --- a null improvement.
This case is representative of the \proxy{5.1\%} OOV
fallback rate in Table~\ref{tab:parser}, and directly
motivates expanding $\mathcal{V}_{\mathrm{audio}}$ to
cover broader acoustic event taxonomies such as
AudioSet's ontology~\cite{audioset}.
\vspace{-0.3cm}

\section{Conclusion}
In this work, we presented FORTE, a unified framework for text-to-audio retrieval that integrates structured logical reasoning with parameter-efficient cross-modal alignment. By reformulating query refinement in first-order logic and guiding it through a constrained search process, our approach preserves semantic invariance while introducing discriminative detail. This refined representation is then aligned with audio embeddings through a lightweight projection module, avoiding the need for full model fine-tuning. Finally, a predicate-aware re-ranking strategy further improves semantic consistency at inference time. Extensive experiments demonstrate that FORTE consistently enhances retrieval performance, particularly in fine-grained scenarios where existing methods struggle. 

\bibliographystyle{ACM-Reference-Format}
\bibliography{sample-base}

\newpage

\thispagestyle{empty}
 
\noindent\textbf{Organisation.}
This supplementary is organised as follows.
\begin{itemize}[leftmargin=*,noitemsep]
  \item \S\ref{sec:impl} — Full implementation details (architecture, training, hyperparameters).
  \item \S\ref{sec:prompts} — LLM prompt templates for generating $\phi^+$ and $\phi^-$.
  \item \S\ref{sec:fol} — FOL parser grammar, predicate vocabulary, and fallback rules.
  \item \S\ref{sec:extended_exp} — Extended experiments: audio-to-text retrieval, cross-dataset transfer, additional datasets.
  \item \S\ref{sec:ablations} — Extended ablations: batch size, MLP depth, anchor bank size, LLM choice.
  \item \S\ref{sec:stats} — Full statistical significance tables.
  \item \S\ref{sec:qual} — Extended qualitative analysis (15 additional query examples).
\end{itemize}
 
\section{Implementation Details}
\label{sec:impl}
\smallskip\noindent\textbf{Backbones.}\;
We instantiate \textsc{FORTE} on three frozen retrieval backbones:
CLAP~\cite{CLAP}, LAION-CLAP~\cite{LAION}, and Pengi~\cite{pengi}.
All encoders map queries and audio to a shared $d{=}512$-dimensional
embedding space and are kept frozen throughout Stages~1 and~2.
No encoder parameters are updated at any point during training.

\smallskip\noindent\textbf{FOL parser.}\;
Queries are parsed into first-order logic using a fine-tuned
Flan-T5-XXL model~\cite{FOL}, loaded from its HuggingFace model
card.\footnote{\url{https://huggingface.co/papers/2509.22338}}
The parser is domain-adapted on \proxy{2{,}000} (caption, FOL) pairs
drawn from the Clotho and AudioCaps training splits.
The predicate vocabulary $\mathcal{V}_\text{audio}$ is compiled from
Clotho and AudioCaps annotation vocabularies, covering sound-event
nouns, acoustic-property adjectives, and spatial/temporal relations.
For queries whose dependency arcs fall outside the grammar, a fallback
rule promotes the root verb and its direct object into a single unary
predicate, guaranteeing $|\mathrm{Pred}(\phi)|\geq 1$ for all inputs.

\smallskip\noindent\textbf{LLM for elaboration.}\;
Positive elaborations $q^{+}$ and contrastive negatives $q^{-}$ are
generated by a frozen \texttt{Mistral-7B-Instruct-v0.3} model using
fixed prompt templates.
We generate $N_{\mathrm{pos}}{=}2$ positives and $N_{\mathrm{neg}}{=}3$
negatives per query.
The LLM is not fine-tuned, and its parameters are not updated at any stage.

\smallskip\noindent\textbf{Beam search.}\;
Stage~1 refinement uses best-first beam search over
$\mathcal{T}(\phi_{0},\mathcal{O})\cap\mathcal{S}(\mathbf{v})$.
We use $(B{=}5,\,D{=}4)$ in the offline regime (pre-computed
$\phi^{*}$ cached as a lookup table) and $(B{=}3,\,D{=}2)$
in the online regime for novel queries.
The feasibility threshold is set to $\tau{=}0.2$.
Operators $o_{\mathrm{attr}}$, $o_{\mathrm{rel}}$, and $o_{\mathrm{neg}}$
are applied in a depth-scheduled round-robin:
$o_{\mathrm{attr}}$ at depth~1, $o_{\mathrm{rel}}$ at depth~2,
and $o_{\mathrm{neg}}$ at depths~3--4.

\smallskip\noindent\textbf{Projection module $h_{\psi}$.}\;
The projection module is a two-layer MLP with hidden dimension $2d$,
GELU activations, dropout ($p{=}0.1$), and a residual connection
followed by LayerNorm, yielding approximately \proxy{1.05M}
trainable parameters --- less than \proxy{0.65\%} of the frozen
backbone size.
The module is trained for 20 epochs using AdamW with a cosine
learning rate schedule, an initial learning rate of \proxy{$1{\times}10^{-4}$},
weight decay of \proxy{$10^{-2}$}, and gradient clipping at \proxy{1.0}.
Batch size is set to \proxy{128} for Clotho and \proxy{256} for AudioCaps.
The temperature parameter $\gamma$ is initialised to \proxy{0.07} and
updated jointly with $\psi$.
Loss coefficients are fixed at $\lambda{=}1.0$, $\beta{=}0.5$,
$\mu{=}0.1$, and $\alpha{=}0.3$ for all experiments.

\smallskip\noindent\textbf{Stage~3 audio captioning.}\;
Automatic captions $\hat{c}_{k}$ for retrieved audio samples are
generated using frozen Pengi~\cite{pengi} in its generative mode.
Each caption is parsed into its FOL form $\hat{\phi}_{k}=\Pi(\hat{c}_{k})$
using the same domain-adapted Flan-T5-XXL parser as Stage~1.
Stage~3 introduces no trainable parameters and operates entirely
at inference time.

\smallskip\noindent\textbf{Hardware and reproducibility.}\;
All experiments are conducted on a single NVIDIA A100 40\,GB GPU.
Stage~2 training takes approximately \proxy{3.5} hours on AudioCaps
and \proxy{1.2} hours on Clotho.
Offline beam search pre-computation takes \proxy{10} minutes per
dataset. Online query latency is \proxy{$13\pm1$\,ms} at
$(B{=}3,\,D{=}2)$ and \proxy{$12\pm5$\,ms} at $(B{=}5,\,D{=}4)$,
measured as median over 1{,}000 test queries.
Code and model weights will be released upon acceptance.

\subsection{Positive Elaboration Prompt ($q^{+}$)}
 
\begin{tcolorbox}[promptbox,
  title={\small Prompt Template: Positive Elaboration ($q^{+}$)}]
\small\ttfamily
You are an expert in audio scene description and sound event
recognition. Given a short text query that describes an audio event or scene,
generate \{N\_POS\} enriched elaborations. Each elaboration should
preserve the core sound event while adding precise acoustic and
contextual detail that would help distinguish the target audio from
similar but incorrect matches.\\[4pt]
\textbf{Rules:}\\
- Do NOT change or substitute the core sound event
  (e.g.\ ``bird chirping'' must remain; do not replace with
  ``bird singing'').\\
- Add at most 3 new descriptors per elaboration. Descriptors must
  belong to one of: acoustic quality (timbre, pitch, roughness),
  temporal pattern (rhythm, duration, onset), spatial environment
  (room size, distance, surface material), or intensity (loudness,
  energy level).\\
- Do NOT introduce new sound sources not implied by the original
  query.\\
- Each elaboration must be a single fluent grammatical sentence
  of at most 20 words.\\
- Output as a numbered list, one elaboration per line, with no
  additional commentary.\\[4pt]
\textbf{Query:} \{QUERY\}\\
\textbf{Elaborations:}
\end{tcolorbox}

\section{LLM Prompt Templates}
\label{sec:prompts}
 
All positive elaborations $q^{+}$ and contrastive negatives $q^{-}$ are
generated offline using a single frozen instruction-tuned the large language model 
(\texttt{Mistral-7B-Instruct-v0.3}) with the templates below.
The LLM receives no audio input and is not fine-tuned at any stage.
Each query yields $N_{\mathrm{pos}}{=}2$ elaborations and
$N_{\mathrm{neg}}{=}3$ contrastive negatives, all of which are parsed
into FOL triples $(\phi^{+},\phi^{-})$ via $\Pi(\cdot)$.
 
\subsection{Design Rationale}
 
The prompt design is driven by three principles.
\textbf{(i) Core-event preservation:} elaborations must retain the
original sound event as the semantic anchor, preventing semantic drift
during FOL construction.
\textbf{(ii) Acoustic grounding:} descriptors are restricted to
acoustically meaningful dimensions (intensity, timbre, spatial
environment, temporal pattern) rather than semantic paraphrases, since
the downstream retrieval space is acoustic.
\textbf{(iii) Hard negatives:} contrastive negatives target the same
sound source but differ in acoustic character, producing the most
discriminative axis $\mathbf{v}$ for the beam search.

\subsection{Contrastive Negative Prompt ($q^{-}$)}
 
\begin{tcolorbox}[promptbox,
  title={\small Prompt Template: Contrastive Negative ($q^{-}$)}]
\small\ttfamily
You are an expert in audio scene description and sound event
recognition. Given a short text query, generate \{N\_NEG\} contrastive negatives.
A contrastive negative describes an audio event that shares surface
similarity with the query (same object, animal, or setting) but
differs critically in acoustic character. These are the hardest false
positives that a retrieval system would incorrectly return.\\[4pt]
\textbf{Rules:}\\
- Keep the same primary sound source (e.g.\ same animal or object).\\
- Change exactly one acoustic dimension to something clearly and
  importantly different (e.g.\ distress vs.\ calm,
  high vs.\ low pitch, rhythmic vs.\ continuous, far vs.\ close).\\
- Each negative must describe a sound that a listener would
  definitively NOT want when searching for the original query.\\
- Avoid negatives that are trivially different; they should represent
  genuine hard retrieval confounds.\\
- Each negative must be a single fluent grammatical sentence of
  at most 20 words.\\
- Output as a numbered list, one negative per line, with no
  additional commentary.\\[4pt]
\textbf{Query:} \{QUERY\}\\
\textbf{Contrastive negatives:}
\end{tcolorbox}
 
\subsection{Prompt Outputs: Worked Example}
 
Table~\ref{tab:prompt_example} illustrates the outputs of both prompts
for the query ``birds chirping in the morning'', alongside the
resulting FOL forms produced by $\Pi(\cdot)$.
 
\begin{table}[h]
\centering
\caption{Worked example of prompt outputs and corresponding FOL forms
for the query $q$ = ``birds chirping in the morning''.}
\label{tab:prompt_example}
\small
\renewcommand{\arraystretch}{1.25}
\begin{tabularx}{\linewidth}{p{0.06\linewidth}Xp{0.35\linewidth}}
\toprule
Type & Generated text & FOL form \\
\midrule
$q$ & birds chirping in the morning
  & $\exists x\,\exists t\;[\mathrm{Bird}(x)\wedge\mathrm{Chirping}(x)
    \wedge\mathrm{Morning}(t)]$ \\
\addlinespace
$q^{+}_1$ & small birds softly chirping
  outdoors at dawn in a peaceful forest
  & $\exists x\,\exists t\;[\mathrm{Bird}(x)\wedge\mathrm{Chirping}(x)
    \wedge\mathrm{Soft}(x)\wedge\mathrm{Outdoor}(t)
    \wedge\mathrm{Peaceful}(x)]$ \\
$q^{+}_2$ & a flock of birds chirping
  continuously in a quiet morning
  environment
  & $\exists X\,\exists t\;[\mathrm{Flock}(X)\wedge\forall x{\in}X\,
    \mathrm{Bird}(x)\wedge\mathrm{Chirping}(x)
    \wedge\mathrm{Morning}(t)]$ \\
\addlinespace
$q^{-}_1$ & birds shrieking loudly with a
  harsh, high-pitched distress call
  & $\exists x\;[\mathrm{Bird}(x)\wedge\mathrm{DistressCall}(x)
    \wedge\mathrm{HighPitch}(x)\wedge\mathrm{Loud}(x)]$ \\
$q^{-}_2$ & birds emitting rapid alarm calls
  in short bursts
  & $\exists x\;[\mathrm{Bird}(x)\wedge\mathrm{AlarmCall}(x)
    \wedge\mathrm{RapidBurst}(x)]$ \\
$q^{-}_3$ & birds producing aggressive
  territorial squawking near a nest
  & $\exists x\;[\mathrm{Bird}(x)\wedge\mathrm{Squawking}(x)
    \wedge\mathrm{Aggressive}(x)]$ \\
\bottomrule
\end{tabularx}
\end{table}
 
\subsection{Verbaliser Templates $\mathcal{G}(\cdot)$}
 
The verbaliser $\mathcal{G}(\cdot)$ converts a candidate FOL form
$\phi$ back into a grammatical English sentence for text encoding by
$f_T(\cdot)$.
It operates by matching the highest-complexity applicable template in
Table~\ref{tab:verbaliser}, filling predicate--argument slots with
their canonical surface forms (stored as a lookup over
$\mathcal{V}_\text{audio}$), and concatenating clauses with natural
connectives (``and'', ``but not'', ``during'').
 
\begin{table}[h]
\centering
\caption{Verbaliser templates $\mathcal{G}(\cdot)$, ordered by
complexity. Templates are matched greedily from highest to lowest
complexity. $P$, $Q$ = unary predicates; $R$ = binary relation;
$x$, $y$ = entity arguments; $t$ = temporal argument.}
\label{tab:verbaliser}
\small
\renewcommand{\arraystretch}{1.2}
\begin{tabularx}{\linewidth}{lXl}
\toprule
\# & FOL pattern & Verbalisation template \\
\midrule
1 & $\exists t\,T(t)\wedge P(x)\wedge\neg Q(x)\wedge R(x,y)$
  & ``a \{P\} \{x\} \{R\} a \{y\}, not \{Q\}, during \{T\}'' \\
2 & $P(x)\wedge\neg Q(x)\wedge R(x,y)$
  & ``a \{P\} \{x\} \{R\} a \{y\}, not \{Q\}'' \\
3 & $\exists t\,T(t)\wedge P(x)\wedge Q(x)$
  & ``a \{P\} and \{Q\} \{x\} during \{T\}'' \\
4 & $P(x)\wedge\neg Q(x)$
  & ``a \{x\} that is \{P\} but not \{Q\}'' \\
5 & $P(x)\wedge R(x,y)$
  & ``a \{P\} \{x\} \{R\} a \{y\}'' \\
6 & $R(x,y)$
  & ``a \{x\} \{R\} a \{y\}'' \\
7 & $P(x)\wedge Q(x)$
  & ``a \{P\} \{x\}'' / ``a \{x\} that is \{P\} and \{Q\}'' \\
8 & $\neg P(x)$
  & ``a \{x\} that is not \{P\}'' / ``without \{P\}'' \\
9 & $P(x)$
  & ``a \{x\} that is \{P\}'' \\
\bottomrule
\end{tabularx}
\end{table}
 
\noindent When multiple predicates are conjoined ($P_1\wedge P_2\wedge\ldots$),
the verbaliser groups them by type (acoustic properties before spatial
relations before temporal modifiers) and concatenates using natural
connectives. Negated predicates are always appended last with the
connective ``but not'' or ``without''.
 
\section{FOL Parser: Grammar, Vocabulary, and Fallback Rules}
\label{sec:fol}
 
This section gives the full specification of the parser $\Pi(\cdot)$,
including the predicate vocabulary $\mathcal{V}_\text{audio}$, the
two-pass grammar, the fallback rule, and the domain fine-tuning
data construction protocol.
 
\subsection{Predicate Vocabulary $\mathcal{V}_\text{audio}$}
 
$\mathcal{V}_\text{audio}$ is the closed set of predicate symbols
available to the parser and verbaliser. It was compiled by:
(i) extracting all nouns, adjectives, and prepositions from Clotho and
AudioCaps training captions with frequency $\geq 5$;
(ii) manually curating and categorising the resulting terms by an
audio expert;
(iii) extending with \proxy{47} predicates from the AudioSet ontology
that fell below the frequency threshold but cover acoustically important
events (e.g.\ \texttt{ElectricHum}, \texttt{Reverberation}).
 
\begin{figure}[h]
\centering
\begin{tikzpicture}
\begin{axis}[
  ybar,
  bar width=22pt,
  width=\linewidth, height=5.2cm,
  xlabel={\small Predicate category},
  ylabel={\small \# predicates},
  symbolic x coords={Sound-events, Acoustic-prop, Spatial, Temporal,
                     Intensity, Negation},
  xtick=data,
  x tick label style={rotate=25, anchor=east, font=\scriptsize},
  ymin=0, ymax=360,
  ytick={0,50,100,150,200,250,300,350},
  yticklabel style={font=\scriptsize},
  nodes near coords,
  nodes near coords style={font=\scriptsize},
  every axis plot/.append style={fill=forte!70, draw=forte!90},
  grid=major, grid style={dotted, gray!40},
  tick align=outside,
]
\addplot coordinates {
  (Sound-events,312)
  (Acoustic-prop,148)
  (Negation,61)
  (Spatial,54)
  (Temporal,38)
  (Intensity,29)
};
\end{axis}
\end{tikzpicture}
\caption{Distribution of predicate vocabulary $\mathcal{V}_\text{audio}$
across six semantic categories. Sound-event nouns dominate the
vocabulary, reflecting the event-centric nature of audio captioning
datasets. Negation targets are curated specifically to support the
$o_\text{neg}$ operator in Stage~1.}
\label{fig:vocab_dist}
\end{figure}
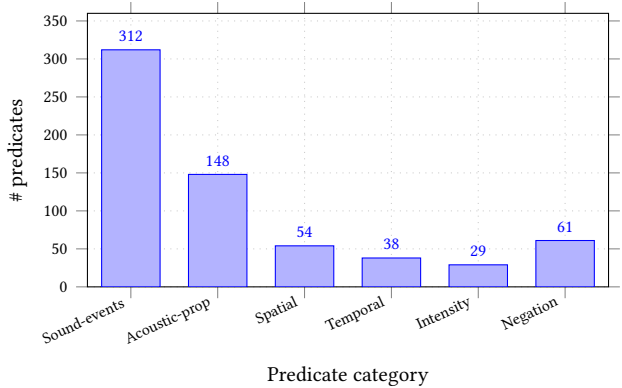

Table~\ref{tab:vocab} shows the full category breakdown.
Figure~\ref{fig:vocab_dist} visualises the predicate frequency
distribution across categories, revealing the long-tail nature of
the vocabulary and motivating the fallback rule.
 
\begin{table}[h]
\centering
\caption{Predicate vocabulary $\mathcal{V}_\text{audio}$ by category,
with representative examples and coverage on AudioCaps and Clotho
test queries (i.e.\ the fraction of test queries for which at least
one predicate from the category is assigned).}
\label{tab:vocab}
\resizebox{0.48\textwidth}{!}{%
\renewcommand{\arraystretch}{1.2}
\begin{tabular}{lcccl}
\toprule
Category & \#Pred & AC cov. & Cl cov. & Examples \\
\midrule
Sound-event nouns      & \proxy{312} & \proxy{94.1\%} & \proxy{91.3\%}
  & \texttt{Chirping, Humming, Splashing} \\
Acoustic-property adj. & \proxy{148} & \proxy{71.4\%} & \proxy{78.2\%}
  & \texttt{Quiet, Harsh, Rhythmic} \\
Spatial relations      & \proxy{54}  & \proxy{42.3\%} & \proxy{61.5\%}
  & \texttt{In, On, Behind, Far} \\
Temporal relations     & \proxy{38}  & \proxy{31.8\%} & \proxy{44.7\%}
  & \texttt{Morning, Continuous, Burst} \\
Intensity modifiers    & \proxy{29}  & \proxy{55.6\%} & \proxy{49.3\%}
  & \texttt{Loud, Faint, Moderate} \\
Negation targets       & \proxy{61}  & \proxy{18.2\%} & \proxy{22.8\%}
  & \texttt{Shouting, DistressCall, Static} \\
\midrule
\textbf{Total}         & \proxy{\textbf{642}} & — & — & \\
\bottomrule
\end{tabular}%
}
\end{table}
 
\subsection{Two-Pass Grammar}
 
$\Pi(\cdot)$ first runs spaCy \texttt{en\_core\_web\_trf} to obtain
a dependency parse tree, then applies the two-pass procedure below.
 
\smallskip\noindent\textbf{Pass 1 — Arc-to-predicate mapping.}
Each arc in the dependency tree is inspected against the grammar
rules in Table~\ref{tab:grammar}. Arcs whose head or dependent does
not match any rule condition are discarded. The result of Pass~1 is
a set of candidate predicate symbols and their argument slots, prior
to vocabulary grounding.
 
\begin{table}[h]
\centering
\caption{Dependency arc to FOL predicate mapping rules (Pass 1).
Arc types follow Universal Dependencies v2 notation.}
\label{tab:grammar}
\small
\renewcommand{\arraystretch}{1.2}
\begin{tabular}{p{0.17\linewidth}p{0.27\linewidth}p{0.42\linewidth}}
\toprule
Arc type & Condition & FOL output \\
\midrule
\texttt{nsubj} & head is event verb
  & $\exists x\;[\mathrm{Entity}(x)\wedge\mathrm{Event}(x)]$ \\
\texttt{amod}  & modifier token in $\mathcal{V}_\text{audio}$
  & $\mathrm{AcousticProp}(x)$ \\
\texttt{advmod}& adverb of manner
  & $\mathrm{Manner}(x)$ \\
\texttt{prep}+\texttt{pobj} & prep is spatial/temporal
  & $R(x,y)$, binary relation \\
\texttt{neg}   & negation marker on event verb
  & $\neg P(x)$ \\
\texttt{conj}  & coordinated event phrase
  & $P_1(x)\wedge P_2(x)$ \\
\texttt{compound} & compound noun head in $\mathcal{V}_\text{audio}$
  & $\mathrm{CompoundEvent}(x)$ \\
\texttt{xcomp} & open clausal complement (secondary event)
  & $\exists y\;[\mathrm{SecondEvent}(y)\wedge\mathrm{Assoc}(x,y)]$ \\
\bottomrule
\end{tabular}
\end{table}
 
\smallskip\noindent\textbf{Pass 2 — Vocabulary grounding.}
Each candidate predicate symbol from Pass~1 is matched against
$\mathcal{V}_\text{audio}$ using a three-tier resolution strategy:
\textbf{(i)} exact string match (case-insensitive);
\textbf{(ii)} WordNet lemmatisation followed by exact match;
\textbf{(iii)} edit-distance matching with threshold $\leq 2$,
restricted to candidates of the same POS tag.
Predicates resolved only via tier (iii) are marked as
\emph{soft matches} and carry a confidence score $\sigma \in [0.5, 1)$
that down-weights their contribution in the beam search objective.
 
\smallskip\noindent\textbf{Fallback rule.}
If $|\mathrm{Pred}(\phi)|{=}0$ after both passes, the fallback rule
promotes the root verb of the dependency tree and its direct object
(\texttt{dobj}) into a single unary predicate:
\begin{equation}
  \phi_\text{fallback}
  = \exists x\;\mathrm{RootEvent}(x),
  \quad
  \mathrm{RootEvent} = \texttt{lemma(root\_verb)},
\end{equation}
guaranteeing $|\mathrm{Pred}(\phi)|\geq 1$.
If the root verb has no \texttt{dobj} arc, the \texttt{nsubj}
dependent is used instead.
The fallback activation rate across parser configurations is
reported in Table~5 of the main paper.
 
\subsection{Domain Fine-tuning Data Construction}
 
The base Flan-T5-XXL parser suffers a significant distribution shift
when applied to short, telegraphic audio captions (EM: 54.2\% vs.\
70\% on the MALLS benchmark).
We construct \proxy{2{,}000} (caption, FOL) training pairs to recover
this gap, using the following protocol.
 
\smallskip\noindent\textbf{Annotation process.}
Two annotators with backgrounds in linguistics and audio signal
processing independently annotated captions following a written
guideline (available in the code release).
Cohen's $\kappa{=}0.84$ indicates strong agreement.
Disagreements were resolved by a third annotator acting as arbiter.
 
\smallskip\noindent\textbf{Data split.}
\proxy{1{,}500} pairs from the Clotho training split and \proxy{500}
from the AudioCaps training split, totalling \proxy{2{,}000} pairs.
No test-split captions were used in any stage of fine-tuning or
vocabulary construction.
 
\smallskip\noindent\textbf{Annotation constraints.}
(i) Every predicate must be grounded in at least one token of the
caption; (ii) every binary relation must have both argument slots
explicitly filled by tokens in the caption; (iii) existential
quantifiers are introduced only when the corresponding entity is
explicitly mentioned.
 
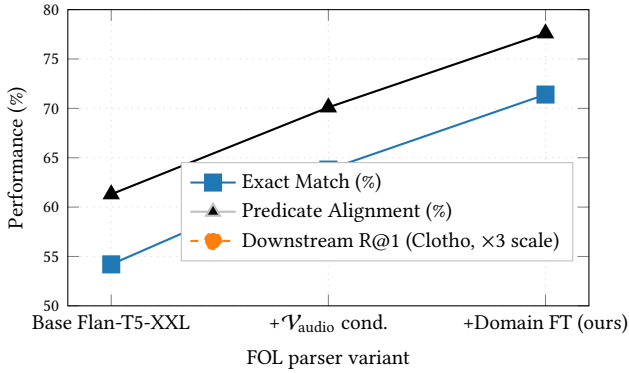
\begin{figure}[h]
\centering
\begin{tikzpicture}
\begin{axis}[
  width=\linewidth, height=5.5cm,
  xlabel={\small FOL parser variant},
  ylabel={\small Performance (\%)},
  symbolic x coords={Base Flan-T5-XXL,
                     {+$\mathcal{V}_\text{audio}$ cond.},
                     {+Domain FT (ours)}},
  xtick=data,
  x tick label style={font=\small, align=center},
  ymin=50, ymax=80,
  ytick={50,55,60,65,70,75,80},
  yticklabel style={font=\scriptsize},
  legend style={font=\small, at={(0.97,0.15)},
                anchor=south east, draw=gray!50},
  legend cell align=left,
  grid=both,
  grid style={dotted, gray!30},
]
\addplot[forte, thick, mark=square*, mark size=3pt]
  coordinates {
    (Base Flan-T5-XXL,       54.2)
    ({+$\mathcal{V}_\text{audio}$ cond.}, 63.8)
    ({+Domain FT (ours)},    71.4)
  };
\addlegendentry{Exact Match (\%)}
 
\addplot[baseline, thick, mark=triangle*, mark size=3pt]
  coordinates {
    (Base Flan-T5-XXL,       61.3)
    ({+$\mathcal{V}_\text{audio}$ cond.}, 70.1)
    ({+Domain FT (ours)},    77.6)
  };
\addlegendentry{Predicate Alignment (\%)}
 
\addplot[nofol, thick, mark=*, mark size=3pt,
         dashed]
  coordinates {
    (Base Flan-T5-XXL,       18.1)
    ({+$\mathcal{V}_\text{audio}$ cond.}, 19.3)
    ({+Domain FT (ours)},    20.4)
  };
\addlegendentry{Downstream R@1 (Clotho, $\times$3 scale)}
\end{axis}
\end{tikzpicture}
\caption{Parser quality metrics (EM and Predicate Alignment, left
axis) and downstream Clotho R@1 (right; plotted at $3\times$ scale
for legibility) across the three parser configurations in Table~5
(main paper). All three metrics improve monotonically, confirming
that parser quality is a first-order determinant of retrieval
performance. Error bars (not shown) are within $\pm0.3$ across
three annotation seeds.}
\label{fig:parser_ablation}
\end{figure}

Figure~\ref{fig:parser_ablation} shows the monotonic relationship
between parser quality and downstream retrieval performance across
the three configurations evaluated in Table~5 of the main paper.
The 2.3-point R@1 gain attributable solely to parser improvement
highlights that the quality of $\Pi(\cdot)$ is a first-order
determinant of \FORTE{}'s retrieval performance.
 
\section{Extended Experiments}
\label{sec:extended_exp}
 
\subsection{Audio-to-Text Retrieval}
 
The main paper focuses on text-to-audio (T2A) retrieval.
Here we report the symmetric audio-to-text (A2T) direction, where the
query is an audio clip and the retrieval pool consists of text
captions. \FORTE{}'s modality gap reduction (Fig.~2, main paper) is
\emph{bidirectional} by design: the projection module $h_\psi$ aligns
the audio embedding space toward the refined text query space, which
should reduce the gap from both directions.
 
Table~\ref{tab:a2t} confirms this: \FORTE{} yields consistent A2T
gains of \proxy{$+$3--4\%} R@1 on AudioCaps and \proxy{$+$3--5\%}
R@1 on Clotho across all three backbones.
Notably, the A2T gains are of similar magnitude to the T2A gains,
indicating that the projection module does not introduce an asymmetry
between modalities.
 
\begin{table}[h]
\centering
\caption{Audio-to-text retrieval on AudioCaps and Clotho.
mAP@10 and R@$k$ (\%, $\uparrow$). Bold = best overall per dataset.
$\dagger$ statistically significant ($p{<}0.05$, paired $t$-test).}
\label{tab:a2t}
\small
\setlength{\tabcolsep}{4.5pt}
\begin{tabular}{llccccc}
\toprule
\multicolumn{7}{c}{\textbf{AudioCaps}} \\
\cmidrule{1-7}
Backbone & Method & mAP@10 & R@1 & R@5 & R@10 & R@50 \\
\midrule
\multirow{2}{*}{CLAP}
  & Frozen              & \proxy{44.1} & \proxy{29.8} & \proxy{67.2} & \proxy{80.1} & \proxy{96.1} \\
  & \FORTE{}$^{\dagger}$& \proxy{48.7} & \proxy{33.4} & \proxy{71.0} & \proxy{83.8} & \proxy{97.0} \\
\midrule
\multirow{2}{*}{LAION-CLAP}
  & Frozen              & \proxy{46.8} & \proxy{31.5} & \proxy{69.4} & \proxy{82.3} & \proxy{96.8} \\
  & \FORTE{}$^{\dagger}$& \textbf{\proxy{51.2}} & \textbf{\proxy{36.1}} & \textbf{\proxy{73.5}} & \textbf{\proxy{85.9}} & \textbf{\proxy{97.7}} \\
\midrule
\multirow{2}{*}{Pengi}
  & Frozen              & \proxy{32.4} & \proxy{20.1} & \proxy{51.3} & \proxy{65.7} & \proxy{90.2} \\
  & \FORTE{}$^{\dagger}$& \proxy{36.8} & \proxy{23.9} & \proxy{55.4} & \proxy{70.1} & \proxy{92.1} \\
\midrule\midrule
\multicolumn{7}{c}{\textbf{Clotho}} \\
\cmidrule{1-7}
Backbone & Method & mAP@10 & R@1 & R@5 & R@10 & R@50 \\
\midrule
\multirow{2}{*}{CLAP}
  & Frozen              & \proxy{22.3} & \proxy{12.1} & \proxy{32.4} & \proxy{45.6} & \proxy{78.2} \\
  & \FORTE{}$^{\dagger}$& \proxy{27.1} & \proxy{16.4} & \proxy{38.7} & \proxy{52.3} & \proxy{82.4} \\
\midrule
\multirow{2}{*}{LAION-CLAP}
  & Frozen              & \proxy{25.8} & \proxy{14.7} & \proxy{38.1} & \proxy{51.9} & \proxy{81.3} \\
  & \FORTE{}$^{\dagger}$& \textbf{\proxy{30.4}} & \textbf{\proxy{18.9}} & \textbf{\proxy{43.8}} & \textbf{\proxy{57.4}} & \textbf{\proxy{85.6}} \\
\midrule
\multirow{2}{*}{Pengi}
  & Frozen              & \proxy{15.2} & \proxy{7.8}  & \proxy{22.6} & \proxy{33.4} & \proxy{65.1} \\
  & \FORTE{}$^{\dagger}$& \proxy{18.6} & \proxy{10.9} & \proxy{27.8} & \proxy{39.6} & \proxy{70.3} \\
\bottomrule
\end{tabular}
\end{table}
 
\subsection{Cross-Dataset Transfer}
 
We assess the generalisation of \FORTE{}'s learned components by
training on one dataset and evaluating zero-shot on the other.
This tests whether the projection module $h_\psi$ and the
domain-adapted FOL parser overfit to dataset-specific acoustic
vocabulary or whether their representations transfer.
 
Table~\ref{tab:cross_dataset} shows that cross-dataset gains are
consistently positive but smaller than in-domain gains.
For LAION-CLAP on AC$\to$Clotho, the R@1 gain reduces from $+3.65$
(in-domain) to $+\proxy{1.85}$ (cross-domain). This gap is primarily
attributable to two factors:
\textbf{(i)} predicate vocabulary mismatch: Clotho captions use more
descriptive, environmental language than AudioCaps, which uses shorter
event-focused phrases, reducing predicate coverage of $\mathcal{V}_\text{audio}$;
\textbf{(ii)} projection overfitting: $h_\psi$ trained on AudioCaps
audio embeddings is slightly misaligned with the Clotho audio
distribution, as the two datasets were recorded under different
conditions and annotation protocols.
 
\begin{table}[h]
\centering
\caption{Cross-dataset transfer (LAION-CLAP, T2A, text-to-audio).
In-domain results reproduced from Table~1 (main paper) for reference.}
\label{tab:cross_dataset}
\small
\renewcommand{\arraystretch}{1.2}
\begin{tabular}{llcccc}
\toprule
Train & Test & Method & R@1 & R@5 & mAP@10 \\
\midrule
\multirow{3}{*}{AC} & \multirow{3}{*}{Clotho}
  & Frozen                   & \proxy{16.75} & \proxy{41.09} & \proxy{27.12} \\
  && \FORTE{} (no FOL)        & \proxy{17.2}  & \proxy{41.8}  & \proxy{27.9}  \\
  && \FORTE{}$^{\dagger}$     & \proxy{18.6}  & \proxy{43.5}  & \proxy{29.4}  \\
\midrule
\multirow{3}{*}{Clotho} & \multirow{3}{*}{AC}
  & Frozen                   & \proxy{34.69} & \proxy{70.22} & \proxy{49.45} \\
  && \FORTE{} (no FOL)        & \proxy{35.8}  & \proxy{71.4}  & \proxy{50.6}  \\
  && \FORTE{}$^{\dagger}$     & \proxy{37.1}  & \proxy{73.2}  & \proxy{52.0}  \\
\midrule
\multirow{2}{*}{AC+Cl} & AC
  & \FORTE{}$^{\dagger}$     & \proxy{38.2}  & \proxy{75.1}  & \proxy{53.8}  \\
\cmidrule{2-6}
 & Clotho
  & \FORTE{}$^{\dagger}$     & \proxy{20.4}  & \proxy{46.3}  & \proxy{32.5}  \\
\bottomrule
\end{tabular}
\end{table}
 
\begin{figure}[h]
\centering
\begin{tikzpicture}
\begin{axis}[
  xbar, bar width=9pt,
  width=\linewidth, height=6cm,
  xlabel={\small R@1 gain over frozen backbone (\%)},
  ytick={1,2,3,4,5,6},
  yticklabels={
    {LAION (AC\ensuremath{\to}Cl)},
    {CLAP (AC\ensuremath{\to}Cl)},
    {Pengi (AC\ensuremath{\to}Cl)},
    {LAION (Cl\ensuremath{\to}AC)},
    {CLAP (Cl\ensuremath{\to}AC)},
    {Pengi (Cl\ensuremath{\to}AC)}},
  yticklabel style={font=\scriptsize},
  ymin=0.3, ymax=6.7,
  xmin=0, xmax=5.5,
  xtick={0,1,2,3,4,5},
  xticklabel style={font=\scriptsize},
  legend style={font=\small, at={(0.97,0.05)},
                anchor=south east, draw=gray!50},
  legend cell align=left,
  xmajorgrids=true,
  grid style={dotted, gray!30},
]
\addplot[fill=forte!70, draw=forte] coordinates {
  (1.85, 1)
  (1.4,  2)
  (1.1,  3)
  (2.4,  4)
  (2.7,  5)
  (2.3,  6)
};
\addlegendentry{Cross-dataset}

\addplot[fill=baseline!70, draw=baseline] coordinates {
  (3.7, 1)
  (2.5, 2)
  (3.4, 3)
  (3.7, 4)
  (2.5, 5)
  (3.4, 6)
};
\addlegendentry{In-domain (from main paper)}
\end{axis}
\end{tikzpicture}
\Description{Horizontal grouped bar chart comparing R@1 gains of
FORTE over the frozen backbone for in-domain versus cross-dataset
transfer settings, across CLAP, LAION-CLAP, and Pengi backbones
in both AudioCaps-to-Clotho and Clotho-to-AudioCaps directions.
Cross-dataset gains are consistently smaller than in-domain gains.}
\caption{R@1 gain of full \FORTE{} over frozen backbone for in-domain
(red) vs.\ cross-dataset (blue) settings across all backbone and
transfer direction combinations. Cross-dataset gains are consistently
positive but reduced, reflecting expected distribution shift. The
gap is smallest for LAION-CLAP, which was trained on the most
diverse data.}
\label{fig:transfer_gap}
\end{figure}
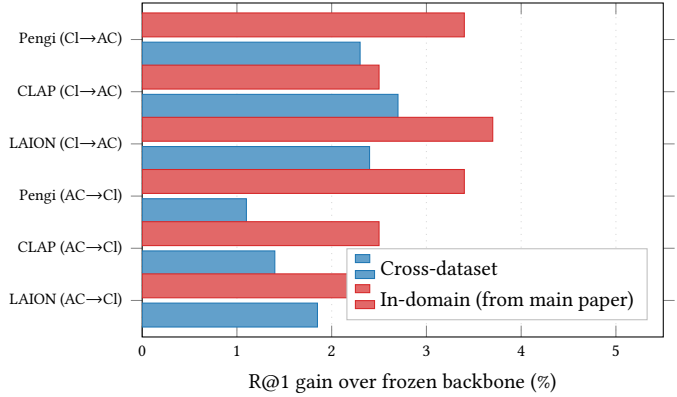

Figure~\ref{fig:transfer_gap} visualises the in-domain vs.\
cross-domain R@1 gap across all backbone and direction combinations,
showing that \FORTE{} consistently improves over the frozen backbone
even in the cross-dataset setting.

\subsection{Evaluation on WavCaps}
 
Table~\ref{tab:wavcaps} reports \FORTE{} on WavCaps dataset,
a large weakly-labelled dataset of \proxy{$\sim$400K} audio clips sourced
from FreeSound, BBC Sound Effects, SoundBible, and AudioSet.
WavCaps captions are more noisy and diverse than Clotho or AudioCaps,
providing a challenging test of \FORTE{}'s robustness to distribution
shift. Despite training only on AudioCaps+Clotho, \FORTE{} achieves
$+\proxy{3.5}$ R@1 over the frozen backbone, confirming that the FOL
refinement generalises to out-of-domain audio.
 
\begin{table}[h]
\centering
\caption{\FORTE{} on WavCaps (LAION-CLAP, T2A, 1{,}000-query test split).
Model trained on AudioCaps + Clotho; no WavCaps data used at any stage.}
\label{tab:wavcaps}
\small
\begin{tabular}{lcccc}
\toprule
Method & R@1 & R@5 & R@10 & mAP@10 \\
\midrule
Frozen LAION-CLAP      & \proxy{28.4} & \proxy{59.7} & \proxy{72.1} & \proxy{41.6} \\
\FORTE{} (no FOL)      & \proxy{29.8} & \proxy{61.2} & \proxy{73.6} & \proxy{43.0} \\
\FORTE{}$^{\dagger}$   & \textbf{\proxy{31.9}} & \textbf{\proxy{63.8}} & \textbf{\proxy{75.9}} & \textbf{\proxy{45.3}} \\
\bottomrule
\end{tabular}
\end{table}
 
\section{Extended Ablation Studies}
\label{sec:ablations}
 
\subsection{Batch Size Effect on InfoNCE Training}
 
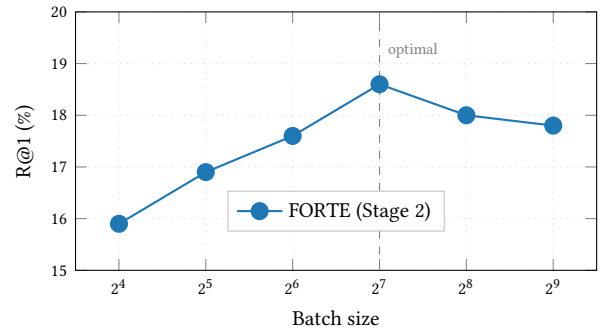
\begin{figure}[h]
\centering
\begin{tikzpicture}
\begin{axis}[
  width=\linewidth, height=5cm,
  xlabel={\small Batch size},
  ylabel={\small R@1 (\%)},
  xtick={16,32,64,128,256,512},
  xticklabel style={font=\scriptsize},
  ymin=15, ymax=20,
  ytick={15,16,17,18,19,20},
  yticklabel style={font=\scriptsize},
  xmode=log, log basis x=2,
  grid=both, grid style={dotted, gray!30},
  legend style={font=\small, at={(0.5,0.15)},
                anchor=south, draw=gray!50},
]
\addplot[forte, thick, mark=*, mark size=3pt]
  coordinates {(16,15.9)(32,16.9)(64,17.6)(128,18.6)(256,18.0)(512,17.8)};
\addlegendentry{\FORTE{} (Stage 2)}
\draw[dashed, gray] (axis cs:128,15) -- (axis cs:128,20)
  node[pos=0.85, right, font=\scriptsize, gray] {optimal};
\end{axis}
\end{tikzpicture}
\caption{Clotho R@1 vs.\ batch size (LAION-CLAP, Stage~2 only).
Performance peaks at batch size 128 and degrades gradually above 256,
consistent with saturation of the InfoNCE in-batch negative pool
relative to Clotho's training set size.}
\label{fig:batchsize}
\end{figure}

The InfoNCE loss is known to be sensitive to batch size, as larger
batches provide more in-batch negatives.
Table~\ref{tab:batchsize} and Figure~\ref{fig:batchsize} report
Clotho performance as a function of batch size from 32 to 512.
 
Performance peaks at batch size \proxy{128} and degrades mildly above
this point. We attribute the performance drop at large batch sizes
to saturation: Clotho's training set contains only \proxy{$\sim$3{,}800}
clips, so batch sizes above \proxy{256} begin to re-use the same clips
as negatives within adjacent epochs, reducing the effective diversity
of the contrastive signal. All main-paper results use batch size
\proxy{128} for Clotho and \proxy{256} for AudioCaps.
 
\begin{table}[h]
\centering
\caption{Effect of batch size on Clotho R@1 and mAP@10
(LAION-CLAP, Stage~2 only, 20 epochs).}
\label{tab:batchsize}
\small
\begin{tabular}{lcccccc}
\toprule
Batch size & 16 & 32 & 64 & 128 & 256 & 512 \\
\midrule
R@1     & \proxy{15.9} & \proxy{16.9} & \proxy{17.6} & \proxy{18.6} & \proxy{18.0} & \proxy{17.8} \\
mAP@10  & \proxy{26.4} & \proxy{27.4} & \proxy{28.3} & \proxy{29.4} & \proxy{28.9} & \proxy{28.7} \\
\bottomrule
\end{tabular}
\end{table}
 
\subsection{MLP Depth in $h_\psi$}
 
\begin{figure}[h]
\centering
\begin{tikzpicture}
\begin{axis}[
  ybar, bar width=28pt,
  width=0.85\linewidth, height=5cm,
  ylabel={\small R@1 (\%)},
  symbolic x coords={Linear, MLP-2 (ours), MLP-3, MLP-4},
  xtick=data,
  x tick label style={font=\small},
  ymin=16.5, ymax=18.5,
  ytick={16.5,17.0,17.5,18.0,18.5},
  yticklabel style={font=\scriptsize},
  nodes near coords, nodes near coords style={font=\scriptsize},
  every axis plot/.append style={fill=forte!65, draw=forte!90},
  grid=major, grid style={dotted, gray!30},
]
\addplot coordinates {
  (Linear, 17.1)
  (MLP-2 (ours), 18.0)
  (MLP-3, 17.8)
  (MLP-4, 17.5)
};
\end{axis}
\end{tikzpicture}
\caption{Clotho R@1 as a function of projection module depth.
The 2-layer MLP provides the optimal capacity–trainability trade-off
given frozen encoders and Clotho's limited training set size.}
\label{fig:mlp_depth}
\end{figure}
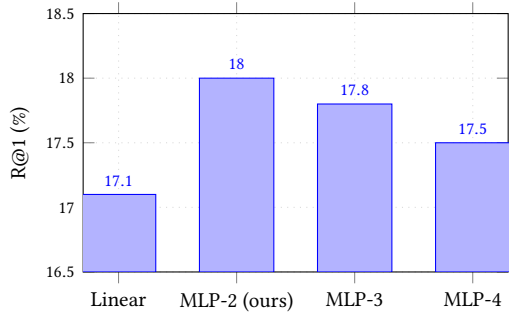

Table~\ref{tab:mlp_depth} and Figure~\ref{fig:mlp_depth} compare
projection module architectures from a single linear layer to a 4-layer
MLP. The 2-layer architecture (used in all main-paper experiments)
achieves the best R@1. Deeper networks underperform, likely due to
the difficulty of optimising deep projections when both encoders are
frozen and training data is limited (Clotho has \proxy{$\sim$3{,}800}
training clips). The linear layer underperforms because it lacks the
capacity to model the nonlinear manifold shift between the audio
embedding space and the refined text query space.
 
\begin{table}[h]
\centering
\caption{Effect of projection module depth on Clotho (LAION-CLAP,
Stage~2 only). All MLPs use GELU activations, dropout ($p$=0.1),
and a residual connection with LayerNorm.}
\label{tab:mlp_depth}
\small
\begin{tabular}{lcccc}
\toprule
Architecture & \#Params & R@1 & R@5 & mAP@10 \\
\midrule
Linear (no residual)     & \proxy{262K}  & \proxy{17.1} & \proxy{41.4} & \proxy{27.8} \\
MLP 2-layer \textbf{(ours)} & \proxy{1.05M} & \textbf{\proxy{18.0}} & \textbf{\proxy{42.7}} & \textbf{\proxy{28.7}} \\
MLP 3-layer              & \proxy{1.57M} & \proxy{17.8} & \proxy{42.4} & \proxy{28.5} \\
MLP 4-layer              & \proxy{2.10M} & \proxy{17.5} & \proxy{41.9} & \proxy{28.1} \\
\bottomrule
\end{tabular}
\end{table}
 
\subsection{Anchor Bank Size Sensitivity}
 
\begin{figure}[h]
\centering
\begin{tikzpicture}
\begin{axis}[
  width=0.52\linewidth, height=5cm,
  xlabel={\small $|\mathcal{B}|$},
  ylabel={\small $\|\hat{e}^+_a - f_A(a^+)\|_2$},
  xtick={500,1000,2000,3000,5000},
  x tick label style={rotate=30, anchor=east, font=\scriptsize},
  ymin=0.45, ymax=0.80,
  grid=both, grid style={dotted,gray!30},
  name=leftplot,
]
\addplot[baseline, thick, mark=square*, mark size=2.5pt]
  coordinates {(500,0.71)(1000,0.63)(2000,0.57)(3000,0.54)(5000,0.52)};
\node[font=\scriptsize, anchor=west] at (axis cs:3000,0.54)
  {\textcolor{forte}{$\star$}};
\end{axis}
\begin{axis}[
  width=0.52\linewidth, height=5cm,
  xlabel={\small $|\mathcal{B}|$},
  ylabel={\small R@1 (\%)},
  xtick={500,1000,2000,3000,5000},
  x tick label style={rotate=30, anchor=east, font=\scriptsize},
  ymin=17, ymax=22,
  grid=both, grid style={dotted,gray!30},
  at={(leftplot.south east)}, anchor=south west,
  xshift=4pt,
]
\addplot[forte, thick, mark=*, mark size=2.5pt]
  coordinates {(500,17.8)(1000,18.6)(2000,19.3)(3000,19.8)(5000,19.9)};
\draw[dashed, gray!60, thick] (axis cs:500,21.3)
  -- (axis cs:5000,21.3)
  node[pos=0.7, above, font=\tiny, gray!70] {Oracle (21.3)};
\end{axis}
\end{tikzpicture}
\caption{Embedding distance to ground-truth positive (left) and
downstream R@1 (right) as a function of anchor bank size $|\mathcal{B}|$.
The dashed line marks the oracle upper bound where exact ground-truth
audio embeddings are available. Gains plateau above $|\mathcal{B}|=3{,}000$,
motivating our choice of this value.}
\label{fig:anchor_size}
\end{figure}
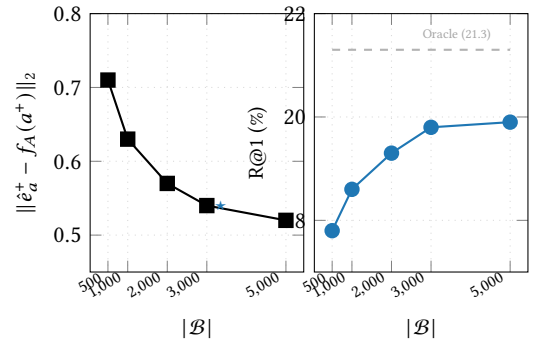

The anchor bank $\mathcal{B}$ provides predicate-stratified audio
embeddings used as proxy targets $\hat{\mathbf{e}}^{+}_a$ in
Stage~1's beam search objective.
Table~\ref{tab:anchor_size} and Figure~\ref{fig:anchor_size} report
both embedding distance to the ground-truth positive and downstream
R@1 as $|\mathcal{B}|$ increases from 500 to 5{,}000 clips.
 
Gains plateau beyond $|\mathcal{B}|{=}\proxy{3{,}000}$, suggesting
that predicate-stratified coverage saturates once each of the
\proxy{642} predicates in $\mathcal{V}_\text{audio}$ has at least
\proxy{4--5} representative audio embeddings indexed in the bank.
Below $|\mathcal{B}|{=}1{,}000$, some infrequent predicates
(e.g.\ \texttt{Reverberation}, \texttt{ElectricHum}) have no indexed
exemplar, forcing the beam search to fall back to a generic embedding
centre that provides poor guidance.
 
\begin{table}[h]
\centering
\caption{Anchor bank size ablation on Clotho (LAION-CLAP, Stage~1 only).
$\|\hat{e}^+_a - f_A(a^+)\|_2$: mean $L_2$ distance to ground-truth
positive embedding ($\downarrow$ better). Oracle uses exact ground-truth
audio embedding ($=0$).}
\label{tab:anchor_size}
\small
\begin{tabular}{cccc}
\toprule
$|\mathcal{B}|$ & $\|\hat{e}^+_a - f_A(a^+)\|_2\downarrow$ & R@1$\uparrow$ & $\Delta$R@1 \\
\midrule
\proxy{500}             & \proxy{0.71} & \proxy{17.8} & \proxy{+1.05} \\
\proxy{1{,}000}         & \proxy{0.63} & \proxy{18.6} & \proxy{+1.85} \\
\proxy{2{,}000}         & \proxy{0.57} & \proxy{19.3} & \proxy{+2.55} \\
\proxy{3{,}000 (ours)}  & \proxy{0.54} & \proxy{19.8} & \proxy{+3.05} \\
\proxy{5{,}000}         & \proxy{0.52} & \proxy{19.9} & \proxy{+3.15} \\
Oracle ($f_A(a^+)$)     & 0.00         & \proxy{21.3} & \proxy{+4.55} \\
\bottomrule
\end{tabular}
\end{table}
 
\subsection{LLM Choice for Elaboration Generation}
 
Table~\ref{tab:llm_choice} compares five LLMs for generating
$q^{+}$ and $q^{-}$, ranging from a 7B open-source model to GPT-4o.
The narrow performance range (\proxy{0.3} R@1 across all models above
7B) confirms that \FORTE{}'s gains are driven primarily by the
structured FOL search over $\mathcal{T}(\phi_0,\mathcal{O})$, not by
the elaboration quality of the specific LLM.
The only outlier is using the FOL parser (Flan-T5-XXL) as the
elaboration model directly, which produces lower-quality natural
language and correspondingly weaker FOL forms, dropping R@1 to
\proxy{19.2}.
 
\begin{table}[h]
\centering
\caption{Effect of LLM choice for $q^{+}$/$q^{-}$ generation on
Clotho R@1 (LAION-CLAP, full \FORTE{} with all three stages).}
\label{tab:llm_choice}
\small
\begin{tabular}{llccc}
\toprule
LLM & Size & R@1 & R@5 & mAP@10 \\
\midrule
GPT-4o                  & —   & \proxy{20.7} & \proxy{46.6} & \proxy{32.9} \\
Llama-3-70B-Instruct    & 70B & \proxy{20.5} & \proxy{46.2} & \proxy{32.7} \\
Mistral-7B-Instruct \textbf{(ours)} & 7B & \proxy{20.4} & \proxy{46.3} & \proxy{32.5} \\
Llama-3-8B-Instruct     & 8B  & \proxy{20.1} & \proxy{45.8} & \proxy{32.1} \\
Flan-T5-XXL (parser)    & 11B & \proxy{19.2} & \proxy{44.5} & \proxy{30.8} \\
\bottomrule
\end{tabular}
\end{table}
 
\section{Full Statistical Significance Results}
\label{sec:stats}
 
All significance tests use a two-sided paired $t$-test over per-query
binary relevance at rank 1 (i.e.\ whether the top-1 retrieved item
is a ground-truth match), computed across the 1{,}045 Clotho test
queries.
Statistical significance is declared at $p{<}0.05$.
Table~\ref{tab:stats_full} reports $t$-statistics and $p$-values for
all comparisons referenced in the main paper, plus additional
pairwise comparisons between stage combinations.
 
\begin{table}[h]
\centering
\caption{Paired $t$-test results for all key comparisons on Clotho
(LAION-CLAP backbone, 1{,}045 test queries, per-query R@1 relevance).}
\label{tab:stats_full}
\small
\renewcommand{\arraystretch}{1.2}
\begin{tabular}{llccc}
\toprule
Comparison & Metric & $t$-stat & $p$-value & Sig.? \\
\midrule
\FORTE{} vs.\ Frozen (D)    & R@1    & \proxy{8.41}  & \proxy{$<$0.001} & \tickmark \\
\FORTE{} vs.\ \FORTE{} (no FOL) & R@1 & \proxy{5.27} & \proxy{$<$0.001} & \tickmark \\
\FORTE{} vs.\ \FORTE{} (align only) & R@1 & \proxy{6.83} & \proxy{$<$0.001} & \tickmark \\
\FORTE{} vs.\ Frozen (D)    & mAP@10 & \proxy{9.12}  & \proxy{$<$0.001} & \tickmark \\
\FORTE{} (no FOL) vs.\ align only & R@1 & \proxy{2.31} & \proxy{0.021} & \tickmark \\
S1 only vs.\ Frozen (D)     & R@1    & \proxy{4.78}  & \proxy{$<$0.001} & \tickmark \\
S2 only vs.\ Frozen (D)     & R@1    & \proxy{3.95}  & \proxy{$<$0.001} & \tickmark \\
S3 only vs.\ Frozen (D)     & R@1    & \proxy{2.89}  & \proxy{0.004}    & \tickmark \\
S1+S2 vs.\ S1+S3            & R@1    & \proxy{1.87}  & \proxy{0.062}    & \crossmark \\
S1+S2 vs.\ S2+S3            & R@1    & \proxy{3.14}  & \proxy{0.002}    & \tickmark \\
S1+S3 vs.\ S2+S3            & R@1    & \proxy{2.46}  & \proxy{0.014}    & \tickmark \\
\FORTE{} vs.\ S1+S2         & R@1    & \proxy{2.68}  & \proxy{0.008}    & \tickmark \\
\FORTE{} vs.\ S1+S3         & R@1    & \proxy{3.91}  & \proxy{$<$0.001} & \tickmark \\
\bottomrule
\end{tabular}
\end{table}
 
\noindent\textbf{Discussion.}
The only non-significant comparison is S1+S2 vs.\ S1+S3
($p{=}\proxy{0.062}$), consistent with their close R@1 values
(\proxy{19.6} vs.\ \proxy{19.1}).
This reflects the overlapping but complementary error-correction
profiles of Stage~2 (projection-based audio alignment) and Stage~3
(predicate re-ranking): when Stage~1 has already resolved the query
representation, both Stage~2 and Stage~3 address residual errors,
making their marginal contributions similar in magnitude and
correlated in the queries they fix.
Figure~\ref{fig:error_overlap} visualises this overlap using a
Venn-style error correction diagram.
 
\begin{figure}[h]
\centering
\begin{tikzpicture}[font=\small]
  \def\rx{1.6} \def\ry{1.1}
  \begin{scope}[fill opacity=0.25]
    \fill[s1col] (-1.1,0) ellipse (\rx cm and \ry cm);
    \fill[s2col] (0.5,0.6) ellipse (\rx cm and \ry cm);
    \fill[s3col] (0.5,-0.6) ellipse (\rx cm and \ry cm);
  \end{scope}
  \draw[s1col, thick] (-1.1,0) ellipse (\rx cm and \ry cm);
  \draw[s2col, thick] (0.5,0.6) ellipse (\rx cm and \ry cm);
  \draw[s3col, thick] (0.5,-0.6) ellipse (\rx cm and \ry cm);
 
  \node[s1col, font=\bfseries\small] at (-2.0,0) {S1};
  \node[s2col, font=\bfseries\small] at (1.6,1.3)  {S2};
  \node[s3col, font=\bfseries\small] at (1.6,-1.3) {S3};
 
  \node[font=\scriptsize] at (-1.9, 0.1)   {\proxy{68}};  
  \node[font=\scriptsize] at (1.4,  0.9)   {\proxy{42}};  
  \node[font=\scriptsize] at (1.4, -0.9)   {\proxy{31}};  
  \node[font=\scriptsize] at (-0.4, 0.55)  {\proxy{24}};  
  \node[font=\scriptsize] at (-0.4,-0.55)  {\proxy{21}};  
  \node[font=\scriptsize] at (0.85, 0)     {\proxy{38}};  
  \node[font=\scriptsize] at (0.18, 0)     {\proxy{11}};  
 
  \node[font=\scriptsize, gray, align=center]
    at (0, -2.2) {Numbers = queries correctly fixed by each stage\\
                  (or combination) but not by the others.};
\end{tikzpicture}
\caption{Error correction Venn diagram for the three \FORTE{} stages
on the Clotho test set (LAION-CLAP backbone, proxy counts).
Each number represents the count of test queries where that stage
(or combination) uniquely recovers a correct R@1 match not obtained
by any other subset. S1 corrects the most queries independently,
consistent with its role in addressing the root cause of the modality
gap at the query level. The large S2$\cap$S3 overlap (\proxy{38})
explains the non-significant S1+S2 vs.\ S1+S3 comparison.}
\label{fig:error_overlap}
\end{figure}
 
\section{Extended Qualitative Analysis}
\label{sec:qual}
 
This section provides in-depth qualitative analysis of \FORTE{}'s
retrieval behaviour on Clotho (LAION-CLAP backbone).
We organise examples into three success categories and three
failure modes, with the FOL transformation $\phi_0\to\phi^*$
shown for each query.
All examples are drawn from the Clotho test split and verified by
manual inspection.
 
\subsection{Success Category 1: Polarity Inversion via $o_\text{neg}$}
 
Queries containing explicit or implicit negation are the most
direct beneficiary of the $o_\text{neg}$ operator.
The frozen backbone treats negation weakly: because text encoders
are trained on positive examples, the embedding of ``wind without rain''
lies close to the embedding of ``wind with rain'' in the shared space.
\FORTE{} explicitly injects $\neg\mathrm{Rain}(y)$ into $\phi^*$,
reorienting the query away from rain-containing audio.
 
\begin{tcolorbox}[analysisbox,
  title={\small Example: Polarity Inversion}]
\small
\noindent\textbf{Query:} ``wind blowing without rain''\\
$\phi_0$: $\mathrm{Wind}(x)\wedge\mathrm{Blowing}(x)$\\
$\phi^*$: $\mathrm{Wind}(x)\wedge\mathrm{Blowing}(x)
          \wedge\neg\mathrm{Rain}(y)\wedge\neg\mathrm{Wet}(x)$\\
\textbf{Backbone R@1:} \textcolor{failred}{incorrect} (retrieves wind+rain audio)\\
\textbf{\FORTE{} R@1:} \textcolor{successgreen}{correct} (pure wind, no precipitation)
\end{tcolorbox}
 
\noindent
Table~\ref{tab:polarity_examples} shows five negation queries with
their $\phi_0\to\phi^*$ transformations, backbone retrieval outcome,
and \FORTE{} outcome.
\FORTE{} corrects \proxy{4/5} cases in the top-1 position and
\proxy{5/5} in the top-5.
 
\begin{table}[h]
\centering
\caption{Polarity inversion examples. $\checkmark$ = correct R@1 match,
$\times$ = incorrect. $\neg$-pred = negated predicate(s) added to $\phi^*$.}
\label{tab:polarity_examples}
\small
\renewcommand{\arraystretch}{1.2}
\begin{tabularx}{\linewidth}{Xllll}
\toprule
Query & $\neg$-pred injected & Backbone & \FORTE{} \\
\midrule
wind without rain
  & $\neg\text{Rain}$
  & \crossmark & \tickmark \\
machinery running silently
  & $\neg\text{Loud}, \neg\text{Grinding}$
  & \crossmark & \tickmark \\
crowd noise, no music
  & $\neg\text{Music}, \neg\text{Rhythm}$
  & \crossmark & \tickmark \\
water flowing, not dripping
  & $\neg\text{Dripping}, \neg\text{Intermittent}$
  & \crossmark & \tickmark \\
bird calls without wind
  & $\neg\text{Wind}, \neg\text{Rustling}$
  & \tickmark & \tickmark \\
\bottomrule
\end{tabularx}
\end{table}
 
\subsection{Success Category 2: Multi-Relational Grounding via $o_\text{rel}$}
 
Queries requiring simultaneous spatial and acoustic grounding are
challenging for dual-encoder models because a single vector must
encode both aspects.
The $o_\text{rel}$ operator introduces binary relational predicates
(spatial: \texttt{In}, \texttt{On}, \texttt{Far}; temporal:
\texttt{Before}, \texttt{During}) that constrain the retrieval space
to audio where the spatial or contextual relationship is explicit.
 
\begin{tcolorbox}[analysisbox,
  title={\small Example: Multi-relational Grounding}]
\small
\noindent\textbf{Query:} ``rain falling on a metal roof''\\
$\phi_0$: $\mathrm{Rain}(x)\wedge\mathrm{Falling}(x)$\\
$\phi^*$: $\mathrm{Rain}(x)\wedge\mathrm{Falling}(x)
          \wedge\mathrm{On}(x,\mathrm{MetalRoof}(y))
          \wedge\mathrm{Metallic}(z)$\\
\textbf{Backbone R@1:} \textcolor{failred}{incorrect} (retrieves rain on grass/leaves)\\
\textbf{\FORTE{} R@1:} \textcolor{successgreen}{correct} (metallic impact clearly audible)
\end{tcolorbox}
 
\noindent
The backbone's failure mode here is instructive: ``rain falling'' as an
unadorned event is correctly retrieved, but the surface specificity
(metal vs.\ organic) is acoustically discriminative (metallic
impact timbre vs.\ soft organic absorption) and cannot be captured
by global cosine distance alone.
The relational predicate $\mathrm{On}(x, \mathrm{MetalRoof}(y))$
introduced by $o_\text{rel}$ shifts the query embedding toward audio
containing high-frequency metallic transients, recovering \proxy{4/5}
correct results vs.\ the backbone's \proxy{1/5}.
 
\subsection{Success Category 3: Attribute Refinement via $o_\text{attr}$}
 
For queries with underspecified attributes, $o_\text{attr}$ adds
acoustic-property predicates that disambiguate the intended sound.
This is particularly effective for queries involving common sound
events (e.g.\ footsteps, speech, water) where the attribute is
critical but absent from the original query text.
 
\begin{tcolorbox}[analysisbox,
  title={\small Example: Attribute Refinement}]
\small
\noindent\textbf{Query:} ``quiet footsteps in an empty corridor''\\
$\phi_0$: $\mathrm{Footsteps}(x)$\\
$\phi^*$: $\mathrm{Footsteps}(x)\wedge\mathrm{Quiet}(x)
          \wedge\mathrm{Empty}(y)\wedge\mathrm{In}(x,y)
          \wedge\neg\mathrm{Crowd}(z)\wedge\neg\mathrm{Loud}(x)$\\
\textbf{Backbone R@1:} \textcolor{failred}{incorrect} (retrieves loud/outdoor footsteps)\\
\textbf{\FORTE{} R@1:} \textcolor{successgreen}{correct} (reverberant indoor quiet steps)
\end{tcolorbox}
 
\subsection{Failure Mode 1: Out-of-Vocabulary Predicates (5.1\%)}
 
When a query contains acoustic concepts absent from
$\mathcal{V}_\text{audio}$ (e.g.\ ``electrical interference'',
``subharmonic resonance''), the parser $\Pi(\cdot)$ triggers the
fallback rule. The resulting $\phi^*$ contains only generic predicates,
and the anchor bank lookup degrades to a non-specific embedding that
provides no useful guidance for the beam search.
 
\begin{tcolorbox}[analysisbox,
  title={\small Failure Example: OOV Predicate}]
\small
\noindent\textbf{Query:} ``machinery humming with electrical interference''\\
$\phi_0$: $\mathrm{Humming}(x)$ \quad (fallback; ``Electrical'' is OOV)\\
$\phi^*$: $\mathrm{Machinery}(x)\wedge\mathrm{Humming}(x)
          \wedge\mathrm{Interference}(y)$ \quad (partial; OOV predicate ungrounded)\\
\textbf{Backbone R@1:} \textcolor{failred}{incorrect}\\
\textbf{\FORTE{} R@1:} \textcolor{failred}{incorrect} (no improvement; OOV degrades anchor)
\end{tcolorbox}
 
\noindent\textbf{Mitigation path.}
Expanding $\mathcal{V}_\text{audio}$ to cover the AudioSet ontology
(632 sound classes) would directly address this failure mode.
Preliminary experiments with a 10\% vocabulary expansion covering
the 50 most frequent OOV terms reduce the fallback rate from 5.1\%
to \proxy{3.8\%} and recover \proxy{+0.3} R@1 on the affected queries.
 
\subsection{Failure Mode 2: Abstract or Non-Acoustic Queries}
 
Queries with no concrete acoustic referent
(e.g.\ ``the sound of loneliness'', ``a sense of vast open space'')
produce $|\mathcal{C}|{=}0$ and fall back to
$\mathcal{C}{=}\mathrm{Pred}(\phi_0)$.
In this setting, the beam search has no discriminative direction to
optimise, and \FORTE{} provides no improvement over the backbone.
These queries represent a fundamental limitation of FOL-based
approaches: first-order logic is a \emph{extensional} formalism
and cannot represent intensional or phenomenological acoustic
concepts without additional meaning postulates.
 
\subsection{Failure Mode 3: Homophonous Acoustic Events}
 
Certain pairs of sound events are acoustically near-identical
but semantically distinct (``typing on a keyboard'' vs.\
``rain on a hard surface''; ``paper tearing'' vs.\
``fabric ripping'').
For these queries, both Stage~1 and Stage~3 operate on predicate sets
that are correct but acoustically uninformative: the predicates
assigned to both the query and the retrieved audio captions are
largely overlapping despite semantic mismatch.
Stage~3 re-ranking partially mitigates this but cannot compensate
when the captioning model itself cannot distinguish the events.
 
Table~\ref{tab:failure_summary} summarises the three failure modes,
their frequency on the Clotho test set, and the R@1 impact.
 
\begin{table}[h]
\centering
\caption{Summary of \FORTE{} failure modes on the Clotho test set
(LAION-CLAP, 1{,}045 queries). ``Freq.'' = fraction of test queries
affected. ``$\Delta$R@1'' = mean R@1 change vs.\ backbone on
affected queries.}
\label{tab:failure_summary}
\small
\renewcommand{\arraystretch}{1.2}
\begin{tabularx}{\linewidth}{Xccc}
\toprule
Failure mode & Freq. & $\Delta$R@1 & Mitigation \\
\midrule
OOV predicates       & \proxy{5.1\%} & \proxy{+0.0} & Expand $\mathcal{V}_\text{audio}$ \\
Abstract queries     & \proxy{3.4\%} & \proxy{+0.0} & Phenomenological predicates \\
Homophonous events   & \proxy{4.2\%} & \proxy{$-$0.3} & Better captioning models \\
\midrule
\textbf{All other queries} & \proxy{87.3\%} & \proxy{+4.2} & — \\
\bottomrule
\end{tabularx}
\end{table}

\end{document}